\documentclass[10pt,journal,letterpaper,compsoc]{IEEEtran}

\usepackage[ruled,vlined]{algorithm2e}
\usepackage{algpseudocode}

\usepackage{amssymb}
\usepackage{array}
\usepackage{amsmath} 
\usepackage{graphicx}
\usepackage{epstopdf}
\usepackage{stmaryrd}
\usepackage{subfig}
\usepackage{paralist}
\usepackage{verbatim}

\usepackage{framed}
\usepackage[dvipsnames]{color}
\definecolor{shadecolor}{gray}{0.85}

\usepackage{soul}

\newif\iflong 
\longfalse

\newif\ifcomm
\commtrue

\newtheorem{thm}{Theorem}
\newtheorem{cor}[thm]{Corollary}

\newcommand{\set}[1]{\left\{#1\right\}}

\newcommand{\unitstep}{\mathbb{U}}
\newcommand{\eps}{\varepsilon}

\newcommand{\argmax}{\arg\max}

\DeclareMathOperator{\maxstep}{maxstep}

\newcommand{\EE}[1]{{\mathbb E}\left[#1\right]}      
\newcommand{\one}[1]{{\mathbb I}_{\{#1\}}}           

\newcommand{\mec}{MEC}
\newcommand{\tsat}{3SAT}
\newcommand{\montsat}{MON3SAT}
\newcommand{\mcgbc}{MC-GBC}
\newcommand{\greedy}{Greedy}

\newcommand{\probrand}{ProbRand}

\newenvironment{proof-sketch}{{\noindent\em Proof Sketch.}\hspace*{0.3em}}{\hfill\QED\medskip}
\newenvironment{proof-of}[1]{{\noindent\em Proof of #1.}\hspace*{0.3em}}{\hfill\qed\medskip}

\newcounter{assumption}
\newcommand{\theassumptionletter}{A}
\renewcommand{\theassumption}{\theassumptionletter\arabic{assumption}}

\newcommand{\MA}{\mathcal{A}}

\newcommand{\MF}{\mathcal{F}}

\newcommand{\MP}{\mathcal{P}}

\newcommand{\GG}{\mathcal{G}}

\newcommand{\beq}{\begin{equation}}
\newcommand{\eeq}{\end{equation}}
\newcommand{\beqa}{\begin{eqnarray}}
\newcommand{\eeqa}{\end{eqnarray}}
\newcommand{\beqan}{\begin{eqnarray*}}
\newcommand{\eeqan}{\end{eqnarray*}}
\newcommand{\ben}{\begin{eqnarray*}}
\newcommand{\een}{\end{eqnarray*}}

\ifcomm
   \newcommand\comm[1]{\textcolor{blue}{ #1}}
\else
   \newcommand\comm[1]{}
   \renewcommand{\todo}[1]{}
\fi

\newcommand{\bm}{\boldmath}

\newcommand{\sss}{\mbox{$\mathbf{s}$}}
\newcommand{\xx}{\mbox{$\mathbf{x}$}}
\newcommand{\pp}{\mbox{$\mathbf{p}$}}

\newcommand{\rr}{\mbox{\bm $r$}}

\newcommand{\yy}{\mbox{$\mathbf y$}}
\newcommand{\G}{\mbox{\bm $G$}}
\renewcommand{\L}{\mbox{\bm $L$}}

\newcommand{\T}{\mbox{\bm $T$}}
\newcommand{\X}{\mbox{\bm $X$}}
\newcommand{\Y}{\mbox{\bm $Y$}}

\newcommand{\MK}{{\cal K}}

\newcommand{\half}{\frac{1}{2}}

\newcommand{\qlica}{{\sc qlICA}}
\newcommand{\bica}{{\sc bICA}}

\makeatletter
\renewcommand\paragraph{\@startsection{paragraph}{4}{\z@}%
    {1.5ex plus .2ex minus .3ex}%
            {-0em}%
                        {\normalsize\bf}}

\newcommand{\captionfonts}{\small}
\long\def\@makecaption#1#2{%
  \vskip\abovecaptionskip
  \sbox\@tempboxa{{\captionfonts #1: #2}}%
  \ifdim \wd\@tempboxa >\hsize
    {\captionfonts #1: #2\par}
  \else
    \hbox to\hsize{\hfil\box\@tempboxa\hfil}%
  \fi
  \vskip\belowcaptionskip}
\makeatother

  {\begin{itemize}%
    \setlength{\itemsep}{0pt}%
    \setlength{\parskip}{0pt}%
        \setlength{\leftmargin}{-10pt}}
  {\end{itemize}}

\begin{document}

\title{On Quality of Monitoring for Multi-channel Wireless Infrastructure Networks}
\author{
Huy Nguyen,~\IEEEmembership{Student Member,~IEEE}, Gabriel Scalosub and Rong Zheng,~\IEEEmembership{Senior Member,~IEEE}
\thanks{H. Nguyen is with the Department of Computer Science, University of Houston, Houston, TX 77204 (e-mail: {\it hanguyen5@uh.edu}).}
\thanks{G. Scalosub is with the Department of Communications System Engineering, Ben-Gurion University of the Negev, Beer-Sheva, Israel (e-mail: {\it sgabriel@bgu.ac.il}).}
\thanks{R. Zheng is with the Department of Computing and Software, McMaster University, Ontario, Canada (e-mail: {\it rzheng@mcmaster.ca}). This work was conducted when the author was with the University of Houston.}
\thanks{An earlier version of this work appeared in the Proceedings of The 11th ACM International Symposium on Mobile Ad Hoc Networking and Computing (Mobihoc), pp. 111--120, 2010.}
}

\IEEEcompsoctitleabstractindextext{%
\begin{abstract}
Passive monitoring utilizing distributed wireless sniffers is an effective
technique to monitor activities in wireless infrastructure networks for fault
diagnosis, resource management and critical path analysis.
In this paper, we introduce a quality of
monitoring (QoM) metric defined by the expected number of active users monitored, and
investigate the problem of maximizing QoM by judiciously assigning sniffers to
channels based on the knowledge of user activities in a multi-channel wireless
network.
Two types of capture models are considered.
The {\it user-centric model} assumes frame-level capturing capability of sniffers
such that the activities of different users can be distinguished while the {\it sniffer-centric model} only utilizes the binary channel
information (active or not) at a sniffer.
For the user-centric model, we show that the implied optimization problem is NP-hard, but a constant
approximation ratio can be attained via polynomial complexity algorithms. For
the sniffer-centric model, we devise stochastic inference schemes to
transform the problem into the user-centric domain, where we are able to apply
our polynomial approximation algorithms. The effectiveness of our proposed
schemes and algorithms is further evaluated using both synthetic data as well as
real-world traces from an operational WLAN.
\end{abstract}

}

\maketitle

\section{Introduction}
\label{sec:introduction}
Deployment and management of wireless infrastructure networks (WiFi, WiMax,
wireless mesh networks) are often hampered by the poor visibility of PHY and
MAC characteristics, and complex interactions at various layers of the protocol
stacks both within a managed network and across multiple administrative
domains.  In addition, today's wireless usage spans a diverse set of QoS
requirements from best-effort data services, to VOIP and streaming
applications. The task of managing the wireless infrastructure is made more
difficult due to the additional constraints posed by QoS sensitive services.
Monitoring the detailed characteristics of an operational wireless  network is
critical to many system administrative tasks including, fault diagnosis,
resource management, and critical path analysis for infrastructure upgrades.

Passive monitoring is a technique where a dedicated set of hardware devices
called {\em sniffers}, or monitors, are used to monitor activities in
wireless networks.  These devices capture transmissions of wireless devices or
activities of interference sources in their vicinity and store the information
in trace files, which can be analyzed distributively or at a central location.
Wireless monitoring~\cite{Yeo04a, Yeo04b,Rodrig05,cheng06,zheng09wiseranalyzer}
has been shown to complement wire side monitoring using SNMP and basestation
logs since it reveals detailed PHY (e.g., signal strength, spectrum density)
and MAC behaviors (e.g, collision, retransmissions), as well as timing
information (e.g., backoff time), which are often essential for wireless
diagnosis. The architecture of a canonical monitoring system consists of three
components:
\begin{inparaenum}[1)]
\item sniffer hardware,
\item sniffer coordination and data collection, and
\item data processing and mining.
\end{inparaenum}

Depending on the type of networks being monitored and hardware capability,
sniffers may have access to different levels of information. For instance,
spectrum analyzers can provide detailed time- and frequency- domain
information.  However, due to the limit of bandwidth or lack of
hardware/software support, it may not be able to decode the captured signal to
obtain frame level information on the fly. Commercial-off-the-shelf network interfaces
such as WiFi cards on the other hand, can only provide frame level
information\footnote{Certain chip sets and device drivers allow inclusion of
header fields to store a few physical layer parameters in the MAC frames.
However, such implementations are generally vendor and driver dependent.}.  The
volume of raw traces in both cases tends to be quite large.
For example, in the
study of the UH campus WLAN, 4 million MAC frames have been collected per sniffer
per channel over an 80-minute period resulting in a total of 8 million distinct
frames from four sniffers.
Furthermore, due to the propagation characteristics
of wireless signals, a single sniffer can only observe activities within its
vicinity.  Observations of sniffers within close proximity over the same frequency band
tend to be highly correlated.  Therefore, two pertinent issues need to be addressed in the design of
passive monitoring systems:
\begin{inparaenum}[1)]
\item what to monitor, and
\item how to coordinate the sniffers to maximize the amount of captured information.
\end{inparaenum}

This paper assumes a generic architecture of passive monitoring systems for
wireless infrastructure networks, which operate over a set of contiguous or
non-contiguous channels or bands\footnote{A channel can be a single frequency
band, a code in CDMA systems, or a hopping sequence in frequency hopping
systems.}.
To address the first question, we consider two categories of capturing models differed
by their information capturing capability. The first category, called the {\it
user-centric model}, assumes availability of frame-level information such that
activities of different users can be distinguished. The second category is the
{\it sniffer-centric model} which only assumes binary information regarding channel
activities, i.e., whether {\it some} user is active in a specific channel near
a sniffer. Clearly, the latter imposes minimum hardware requirements,
and incurs minimum cost for transferring and storing traces. In some cases, due
to hardware constraints (e.g., in wide-band cognitive radio networks) or security/privacy
considerations, decoding of frames to extract user level information is infeasible and thus
only binary sniffer information might be available for surveillance purpose.
We further characterize
theoretically the relationship between the two models.

Ideally, a network administrator would want to perform network monitoring on all
channels simultaneously. However, multi-radio sniffers are known to be large and expensive
to deploy~\cite{jigsaw}. We therefore assume sniffers in our system are low-cost devices
which can only observe one single wireless channel at a time.
To maximize the amount of captured information, we introduce a
quality-of-monitoring (QoM) metric defined as the total expected number of active users
detected, where a user is said to be active at time $t$, if it transmits over one of the wireless channels.
The basic problem underlying all of our models can be cast as {\em finding an assignment of sniffers to channels so as to maximize the QoM}.
QoM is an important metric that quantifies the efficiency of monitoring solutions to systems where it is important to capture as comprehensive information as possible (e.g.: intrusion/anomaly detection~\cite{ids09, IDS03} and diagnosing systems~\cite{diagnosing09, diagnosing08}).

We note that the problem of sniffer assignment, in an attempt to maximize the QoM metric,
is further complicated by the dynamics of real-life systems such as:
\begin{inparaenum}[1)]
\item the user population changes over time (churn),
\item activities of a single user is dynamic, and
\item connectivity between users and sniffers may vary due to changes in channel conditions or mobility.
\end{inparaenum}
These practical considerations reveal the fundamental intertwining of
``learning'', where the usage pattern of wireless resources is to be estimated
online based on captured information, and ``decision making", where sniffer
assignments are made based on available knowledge of the usage pattern.
In fact, in our earlier work~\cite{AroraSZ11}, we prove that during
learning, each instance of the decision making is equivalent to solving an instance of the
sniffer assignment problem with the parameters properly chosen. Thus, effective
and efficient algorithms for the sniffer assignment problem is critical. In
this paper, we focus on designing algorithms that aim at maximizing the QoM
metric with different granularities of {\it a priori} knowledge. The usage
patterns are assumed to be stationary during the decision period.


\paragraph*{Our Contribution}
\label{sec:our-contribution}
In this paper, we make the following contributions toward the design of passive
monitoring systems for multi-channel wireless infrastructure networks

\begin{itemize}
\item We provide a formal model for evaluating the quality of monitoring.
\item We study two categories of monitoring models that differ in the information capturing capability of passive
monitoring systems. For each of these models we provide algorithms and methods
that optimize the quality of monitoring.
\item We unravel interactions between the two monitoring models by devising two methods
to convert the sniffer-centric model to the user-centric domain by exploiting the
stochastic properties of underlying user processes.

\end{itemize}
More specifically, we show that in both the user- and sniffer-centric models
considered, a pure strategy where a sniffer is assigned to a single channel
suffices in order to maximize the QoM. In the {\it user-centric model}, we show
that our problem can be formulated as a covering problem. The problem is proven
to be NP-hard, and constant-approximation polynomial algorithms are provided.
With the {\it sniffer-centric model}, we show that although the only information
retrieved by the sniffers is binary (in terms of channel
activity), the ``structure'' of the underlying
processes is retained and can be recovered. Two different approaches are proposed that utilize the notion of
Independent Component Analysis (ICA)~\cite{Hyvarinen00} and allow mapping the
sniffer assignment problem to the user-centric model.  The first approach,
Quantized Linear ICA (\qlica), estimates the hidden structure by applying
a quantization process on the outcome of the traditional ICA, while the second
approach, Binary ICA (\bica)~\cite{Nguyen2011TSP}, decomposes the
observation data into OR mixtures of hidden components and recovers the
underlying structure.  Finally, an extensive evaluation study is carried out using
both synthetic data as well as real-world traces from an operational WLAN.

The paper is organized as follows. An overview of related work is provided
in Section~\ref{sec:previous-work}. In Section~\ref{sec:problem-formulation},
we formally introduce the QoM metric and the user-centric and sniffer-centric
models for a passive monitoring system.  The NP-hardness and polynomial-time
algorithms for the maximum effort coverage problem that underlies two variants
of the user-centric model are discussed in Section~\ref{sec:user-centric}. The
relationship between the user-centric and sniffer-centric models is established
in Section~\ref{sec:sniffer-centric}, where we also describe two schemes for
solving the QoM problem under the sniffer-centric model. We present the
results of the evaluation study using both synthetic and real traces in
Section~\ref{sec:evaluation}. We discuss issues regarding practical system
implementation in Section~\ref{sec:discuss} and finally conclude the paper in
Section~\ref{sec:conclusion}.

\vspace{0.2cm}

\section{Related Work}
\label{sec:previous-work}
In this section, we provide an overview of related work pertaining to wireless
network monitoring, and binary independent component analysis.

\paragraph*{Wireless monitoring}
There has been much work done on wireless monitoring from a {\em system-level}
approach, in an attempt to design complete systems, and address the
interactions among the components of such systems.  The work in
\cite{Balachandran02,Henderson04} uses AP, SNMP logs, and wired side traces to
analyze WiFi traffic characteristics.  Passive monitoring using multiple
sniffers was first introduced by Yeo {\it et al.} in~\cite{Yeo04a, Yeo04b},
where the authors articulate the advantages and challenges posed by passive
measurement techniques, and discuss a system for performing wireless
monitoring with the help of multiple sniffers, which is based on synchronization and merging of
the traces via broadcast beacon messages. The results obtained for these
systems are mostly experimental.
Rodrig {\it et al.}
in~\cite{Rodrig05}  used sniffers to capture wireless data, and
analyze the performance characteristics of an 802.11 WiFi network. One key
contribution was the introduction of a  finite state machine to infer missing
frames. The Jigsaw system, that was proposed in~\cite{cheng06}, focuses on
large scale monitoring using over 150 sniffers.

A number of recent works focused on the {\em diagnosis of wireless networks to
determine causes of errors}. In~\cite{Chandra06}, Chandra {\it et al.} proposed
WiFiProfiler, a diagnostic tool that utilizes exchange of information among
wireless hosts about their network settings, and the health of network
connectivity. Such shared information allows inference of the root causes of
connectivity problems. Building on their monitoring infrastructure, Jigsaw,
Cheng {\it et al.} \cite{cheng07}  developed a set of techniques for automatic
characterization of outages and service degradation.
They showed how sources of delay at multiple layers (physical through transport) can be reconstructed by using a combination of measurements, inference and modeling. Qiu {\it et al.} in~\cite{Qiu06} proposed a
simulation based approach to determine sources of faults in wireless mesh
networks caused by packet dropping, link congestion, external noise, and MAC
misbehavior.

All the afore-mentioned work focuses on building monitoring infrastructure, and
developing diagnosis techniques for wireless networks. The question of
optimally allocating monitoring resources to maximize captured information
remains largely untouched. In \cite{shin09optimal}, Shin and Bagchi consider
the selection of monitoring nodes and their associated channels for monitoring
wireless mesh networks. The optimal monitoring is formulated as maximum
coverage problem with group budget constraints (denoted \mcgbc), which was
previously studied by Chekuri and Kumar in~\cite{chekuri04maximum}.
The user-centric model results in a problem formulation that is similar to (albeit
different from) the one addressed in~\cite{shin09optimal}. On one hand, we
assume all sniffers may be used for monitoring (hence parting with our problem
being akin to the classical maximum-coverage problem, while on the other hand we focus on the weighted version of the
problem, where elements to be covered have weights. One should note that all
the lower bounds mentioned in~\cite{chekuri04maximum,shin09optimal} do not
apply to our problem.

\paragraph*{Binary independent component analysis}
%
Binary ICA is a special variant of the traditional ICA, where linear mixing of
continuous signals is assumed. In binary ICA,
boolean mixing (e.g., OR, XOR etc.) of binary signals is considered. Existing
solutions to binary ICA mainly differ in their assumptions of prior distribution of
the mixing matrix, noise model, and/or hidden causes. In \cite{Yeredor07},
Yeredor considers binary ICA in XOR mixtures and investigates the identifiability
problem. A deflation algorithm is proposed for source separation based on
entropy minimization. In \cite{Yeredor07} the number of independent random
sources $K$ is assumed to be known. Furthermore, the mixing matrix is a
$K$-by-$K$ invertible matrix.  In \cite{computer06anon-parametric}, an infinite
number of hidden causes following the same Bernoulli distribution is assumed.
Reversible jump Markov chain Monte Carlo and  Gibbs sampler techniques are
applied. In contrast, in our model, the hidden causes may follow different
distributions.  Streith {\it et al.}~\cite{Streich09} study the problem of
multi-assignment clustering for boolean data, where an object is represented by
a boolean attribute vector.
The key assumption made in this work is that elements of the
observation matrix are conditionally independent given the model parameters.
This greatly reduces the computational complexity and makes the scheme amenable
to gradient descent optimization solution; however, the assumption is in
general invalid. In
\cite{Kaban_factorisationand}, the problem of factorization and de-noising of
binary data due to independent continuous sources is considered. The sources
are assumed to be following a beta distribution and not binary.  Finally,
\cite{computer06anon-parametric} considers the under-represented case of less
sensors than sources with continuous noise, while
\cite{Kaban_factorisationand,Streich09} deal with the over-determined case,
where the number of sensors is much larger than the number of sources.

\section{Problem formulation}
\label{sec:problem-formulation}

\subsection{Notation and network model}
\label{sec:network-model}
Consider a system of $m$ sniffers, and $n$ users, where each user $u$ operates
in one of $K$ channels, $c(u) \in \MK=\set{1,\ldots,K}$. The users can be
wireless (mesh) routers, access points or mobile users.  At any point in time,
a sniffer can only monitor packet transmissions over a single channel. We
assume the propagation characteristics of all channels are similar. We
represent the relationship between users and sniffers using an undirected
bi-partite graph $\GG=(S,U,E)$, where $S$ is the set of sniffer nodes and $U$ is
the set of users. Note that $\GG$ represents a general relationship between the users and sniffers, and no propagation or coverage model is assumed. An edge $e=(s,u)\in E$ exists between sniffer $s\in S$ and user $u \in U$ if $s$ can capture the transmission from $u$, or equivalently, $u$ is within the monitoring range of $s$. If transmissions from a
user cannot be captured by any sniffer, the user is excluded from $\GG$. For
every vertex $v \in U\cup S$, we let $N(v)$ denote vertex $v$'s neighbors in
$\GG$. For users, their neighbors are sniffers, and vice versa. We will also
refer to $\G$ as the binary $m \times n$ adjacency matrix of graph $\GG$.

We will consider {\em sniffer assignments} of sniffers to channels, $a:S \rightarrow \MK$.
Given a sniffer assignment $a$, we consider a partitioning of the set of sniffers $S = \bigcup_{k=1}^{K}{S_k}$, where $S_k$ is the set of sniffers assigned to channel $k$.
We further consider the corresponding partition of the set of users
$U = \bigcup_{k=1}^{K}{U_k}$, where $U_k$ is the set of users
operating in channel $k$.
Let $\GG_k=(S_k,U_k,E_k)$ denote the bipartite subgraph of $\GG$ induced by channel $k$. Given any sniffer $s$, we let $N_k(s)=N(s) \cap U_k$, i.e., the set of neighboring users of $s$ that use channel $k$.

A {\em monitoring strategy} determines the channel(s) a sniffer monitors. It
could be a {\it pure strategy}, i.e., the channel a sniffer is assigned to is fixed, or
a {\it mixed strategy} where sniffers choose their assigned channel in each slot
according to a certain distribution.  Formally, let $\MA = \set{a \mid a: S
\rightarrow \MK}$ be the set of all possible assignments.  Let $\pi: \MA
\rightarrow [0,1]$ be a probability distribution over the set of sniffer
assignments. We refer to such a distribution as a mixed strategy.
A pure strategy that selects a single channel per sniffer is a special case of
mixed strategies, namely, $\pi(a) = 1$. It follows that the
pure strategy is generally suboptimal comparing to the mixed strategy. However, as
shown in the next section, the optimal solution can be obtained using just a pure strategy.

In this paper we consider the problem of finding the monitoring strategy that maximizes QoM, defined as the expected number of users detected given the sniffer assignments.
The main notations used in this paper are summarized in Table~\ref{tab:mainnotations}.

\begin{table}[t]
\caption{Notations}
\centering 
\vspace{0.1in}
{\small
\begin{tabular}{c | c}
\hline\hline
$m,n,$ & number of sniffers, users,\\
$K,T$ & channels, and observations\\
\hline
& bi-partite graph representing\\[-0.8ex]
\raisebox{1.2ex}{$\GG$}
& user and sniffer adjacency \\
\hline
$S$ & set of sniffers nodes in $\GG$\\
\hline
$U$ & set of users in $\GG$\\
\hline
$\MA$ & set of all possible sniffer-channel assignments \\
\hline
& vector of $m$ binary random variables\\[-0.8ex]
\raisebox{1.2ex}{$\xx_{m\times 1}$}
& from $m$ sniffers\\
\hline
& vector of $n$ binary random variables\\[-0.8ex]
\raisebox{1.2ex}{$\yy_{n\times 1}$}
& from $n$ users \\
\hline
$\X_{m\times T}$ & collection of $T$ observations of $\xx$ \\
\hline
$\Y_{n\times T}$ & collection of $T$ observations of $\yy$ \\
\hline
$\G_{m \times n}$ & binary adjacency matrix of $\GG$ \\
\hline
$\pp_{1\times n}$ & active probability vector of $n$ users \\
\hline
$c(u)$ & active channel of user $u$ \\
\hline
$\MA(u)$ & sniffer assignments that can monitor user $u$ \\
\hline
$\pi(a)$ & probability distribution of assignment $a$ \\
\hline\hline
\end{tabular}
}
\label{tab:mainnotations}
\end{table}

\subsection{Models for Observing User Access Patterns}
\label{sec:user-access-model}
In this section, two categories of parametric models are proposed to describe the
observability of usage patterns. We assume time is separated into slots,
where each slot represents a fixed duration of time. A user is active if there exists a transmission event from the user during the slot time. In the experiments, slot time is chosen to be on the same order of maximum packet transmission time.
Furthermore, we assume all channel and users' statistics remain stationary for
the monitoring period of $T$ time slots.
\paragraph*{User-centric model}
First, we consider transmission events in the network from the user's
viewpoint.
We assume that $\G$ is known by inspecting the packet header information from each sniffer's captured traces.

In the user-centric model, the transmission probabilities of the users
$\pp=\{p_u|u\in U\}$ are known and assumed to be independent \footnote {
The assumption that user activities are independent has been widely adopted in
literature, examples are \cite{Garba05} and \cite{Kandeepan10}. In the simulation evaluation in Section~\ref{sec:evaluation}, the proposed algorithms are shown to perform well even when such independency is violated.}. $p_u$ denotes
the transmission probability of user $u$.
$p_u$ and $\G$ can be estimated by putting all sniffers in the same channel and iterating through all possible channels for sufficiently long time.
Each user process  may be IID or non-IID over time.

\begin{figure}[t]
\begin{center}
\includegraphics[width=1.5in]{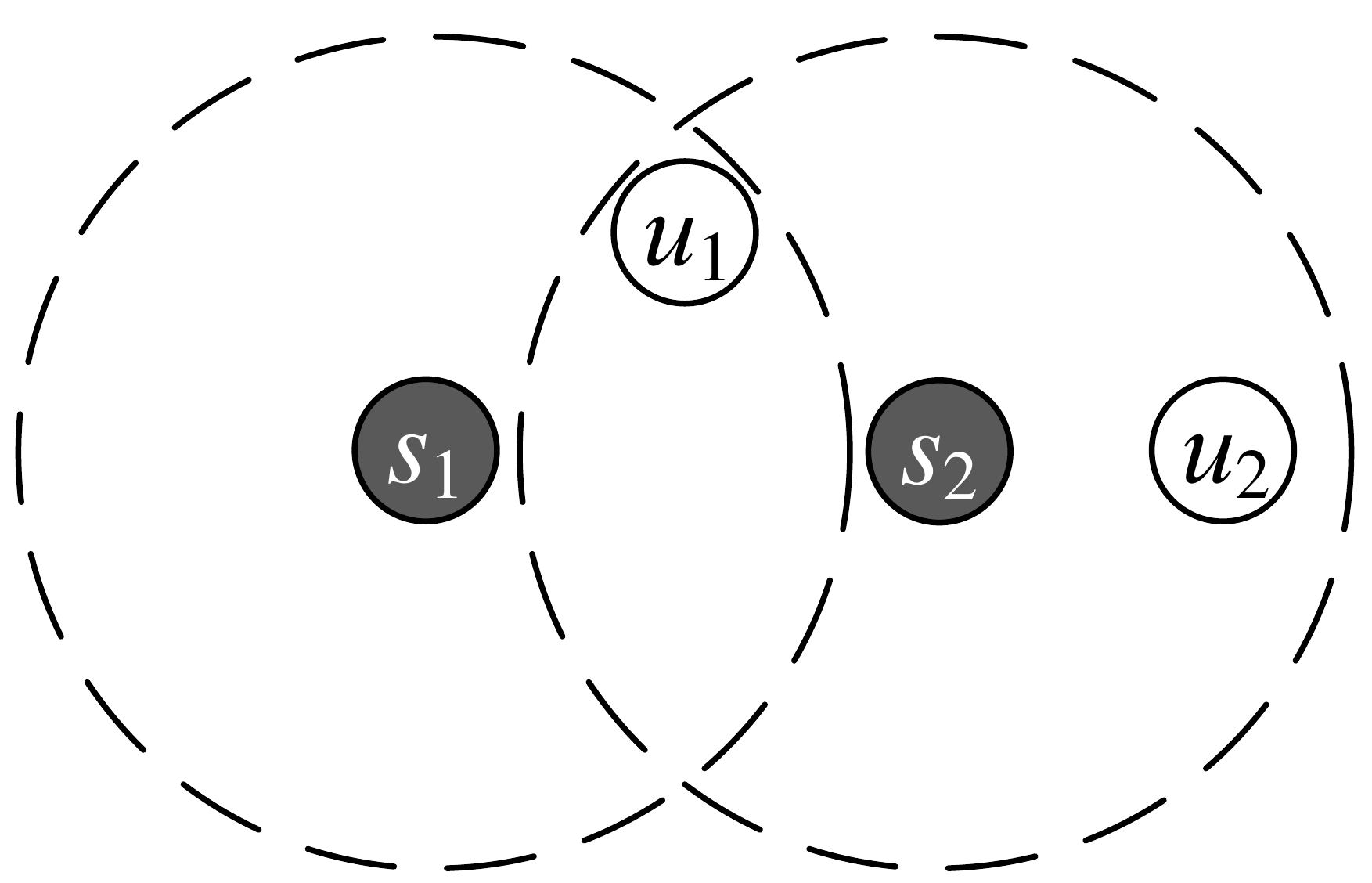}
$$
\G =
\begin{bmatrix}
1&0 \\
1&1 \\
\end{bmatrix}
$$
\caption{A toy example. Users are shown in
white circles and sniffers are shown in black
circles. Sniffer range indicates weather or not a
sniffer can capture a user's transmissions.}
\vspace{-1em}
\label{fig:toy}
\end{center}
\end{figure}

Consider a wireless network with 2 sniffers and 2 users on 2 channels (Figure~\ref{fig:toy}).
User $u_1$ and $u_2$ are active on channels 1 and 2, respectively. Transmission probabilities of users are $p_1 = 0.2$ and $p_2 = 0.5$. User-centric model
assumes $\G$ and $\pp = \{p_1, p_2\}$ are available. Note that the maximum value of QoM in the above network is 0.7 attained when $s_1$ and $s_2$ are assigned to channels 1 and 2, respectively.

\paragraph*{Sniffer-centric model}
The user-centric model requires detailed knowledge of each user's activities.
This necessitates frame-level capturing capability by the passive monitoring
system. In the sniffer-centric model, only {\bf binary} information ({\it on} or {\it
off}) of the channel activity at each sniffer is observed.




We denote by $\xx_k$ the binary vector
of observations when all sniffers operate on channel $k$
and by $\X_k$ the collection of $T$ realizations of $\xx_k$.
We assume that sniffers' observations on different channels are independent. However, dependency exists
among observations of sniffers operating in the same channel (as a result of transmissions made by the same set of users). Given an assignment $a$, a complete characterization of the sniffers' observations is
given by the joint probability distribution $\MP_a(\xx_k)$, $k = 1, \ldots, K$. Here, $\MP_a(\xx_k)$ is  implicitly dependent on the assignment $a$ such that if sniffer $i$ is not assigned to the $k$'th channel, its binary observation $\xx_k(i)$ is always zero.
By independence of different channels we have $\MP_a(\xx) = \prod_{k=1}^{K}{\MP_a(\xx_k)}$.

Consider again the network in Figure 1. Over $T$ time slots, we have two observation
matrices $\X_1$ and $\X_2$ at the same dimension ($2 \times T$) corresponding to the activities on two channels.
The first and second line in each matrix contain observations from sniffers $s_1$ and $s_2$, respectively.
Sniffer-centric model assumes only the availability of $\X_1$ and $\X_2$, while $\G$ and $\pp$ are unknown.

Clearly, the sniffer-centric model is not as expressive as the user-centric
model (formally characterized in Section~\ref{sec:sniffer-centric-vs-user-centric}).
However, it has the advantage of being based on {\em aggregated} statistics, which are
likely to remain stationary in the presence of moderate user-level dynamics, such
as joining and leaving the networks, or  changes in transmission activities
(e.g., busy or thinking time).  Furthermore, obtaining such binary information
is less costly in both hardware requirements and communication/storage
complexity.
\section{QoM under the User-Centric Model}
\label{sec:user-centric}

Under the user-centric model the goal is to maximize the expected number of
active users monitored. Recall that $p_u$ is the transmission probability of
user $u$. This problem can be formulated formally by:
\beq
\begin{array}{ll}
\underset{\pi(a)}{\max} & \sum_{u \in U}{p_u \sum_{a \in \MA(u)}{\pi(a)}} \\
\mbox{s.t.} & \pi(a) \in [0,1]  \\
& \sum_{a \in \MA}{\pi(a)} = 1,
\end{array}
\eeq
where $\MA(u)$ is the set of assignments that monitors
user $u$, i.e., $\MA(u) = \set{a \mid \exists s \in N(u) \mbox{ s.t. } a(s)=c(u)}$.
The objective function calculates the opportunity for all users to be monitored
given the probability of each assignment (QoM). It can be written as,
\beq
\sum_{a \in \MA}{\pi(a) \sum_{u \in U} {p_u\cdot \one{a \in \mathcal{A}(u)}}},
\label{eq:alter_obj}
\eeq
where $\one{\cdot}$ is an indicator function.  From
Eq.~\eqref{eq:alter_obj} it is clear that a pure strategy can be adopted and is optimal,
i.e., an optimal assignment is given by
\beq
a^* = \argmax_a{\sum_{u \in U}{p_u\cdot \one{a \in \mathcal{A}(u)}}}.
\eeq

\subsection{MAX-EFFORT-COVERAGE problem}
Under the user-centric model, the objective to find the sniffer-channel assignment
that can monitor the largest (weighted)
set of users, subject to the constraint that each sniffer can only monitor one of the $K$ channels at a time.
We henceforth refer to the problem as MAX-EFFORT-COVERAGE (\mec) problem.
Note that in \mec\, the
weights can in fact be any non-negative values and are not limited to $[0,1]$.
The \mec\ problem can be cast as the following integer program (IP):

\beq
\label{eq:ip-mec}
\begin{array}{rll}
\underset{z}{\max} & \sum_{u \in U} p_u y_u \\[0.2cm]
\mbox{s.t.}
& \sum_{k=1}^K z_{s,k} \leq 1 & \quad \forall s\in S \\
& y_u \leq  \sum_{s \in N(u)} z_{s,c(u)} & \quad \forall u \in U\\
& y_u \leq 1 & \quad \forall u\in U \\
& y_u, z_{s,k} \in \set{0,1} & \quad \forall u,s,k.
\end{array}
\eeq

Each sniffer is associated with a set of binary decision variables, $z_{s,k} =
1$ if the sniffer is assigned to channel $k$; 0, otherwise.
$y_u$ is a binary variable indicating whether or not user $u$ is monitored, and $p_u$ is the weight associated with user $u$.
The objective function characterizes the number of (weighted) users that can be monitored with assignment $z$.

One should first note that the problem is trivial if $K=1$, since all sniffers would simply be assigned to the sole available channel. We can therefore assume that $K \geq 2$.
The \mec\ problem can be viewed as a special case of the \mcgbc\ (mentioned in Section~\ref{sec:previous-work}), where all sniffers are used.
One should note that previous hardness results for \mcgbc\ (both NP-hardness,
as well as hardness of approximation) were based on a reduction to the standard
maximum coverage problem. It follows that none of these proofs are applicable to the
\mec\ problem. Surprisingly, there has not been any work done explicitly on the
\mec\ problem, which seems to be a natural and important variant of the maximum
coverage problem.

\subsection{Hardness of \mec}
\label{sec:mec-hardness}
In what follows we show that the \mec\ problem is NP-hard for $K\geq 2$, even
for the unweighted case (i.e., where $p_u =1$ for all $u \in U$).
The hardness of the \mec\ problem actually follows from the choices
available to the different sniffers. It is inherently different from the
hardness suggested for the \mcgbc\ problem, which follows from limiting the
number of sniffers one is allowed to use.
We prove hardness of \mec\ using a reduction from the problem of Monotone-3SAT (\montsat), which is known to be NP-hard (see~\cite{gold78complexity,garey79computers}). In \montsat\ we are given as input an instance of \tsat\ where every clause consists of either solely positive variables, or solely negated variables. The goal is to decide whether or not there exists an assignment which satisfies all clauses.


In~\cite{Nguyen2010mobihoc}, we proved that the the unweighted \mec\ problem is NP-hard, even for $K=2$.
The result implies that one would have to settle for approximate solutions to \mec. We first note that Guruswami and Khot show in~\cite{guruswami05hardness} that \montsat\ is NP-hard
to approximate within a factor of $7/8 + \eps$ for every $\eps>0$.
The following is a corollary of the above fact:
\begin{cor}
The \mec\ problem is NP-hard to approximate to within a factor of $7/8+\eps$ for every $\eps>0$.
\end{cor}

\subsection{Algorithms for \mec}
\label{sec:mec-algorithms}
Since \mec\ is a special case of the \mcgbc\ problem, we can use the available approximation algorithms for \mcgbc\ (e.g., \cite{chekuri04maximum,shin09optimal}) to solve our problem in the user-centric model. In what follows we give a brief overview of the algorithms we use.

\paragraph*{The \greedy\ algorithm}
The \greedy\ algorithm iteratively assigns sniffers to users, where at each step it chooses the sniffer and the assignment that (locally) maximizes the weight of coverage of those not yet monitored users.

It is proven in~\cite{chekuri04maximum} that in the unweighted case, i.e., where all users have the same weight, \greedy\ guarantees to produce a $\half$-approximate solution, and that this is tight.
The following theorem shows that the same holds also for the weighted case, which generalizes the \mec\ problem.
\begin{thm}
\label{thm:greedy-approximation-ratio}
\greedy\ is a $\half$-approximation algorithm for the weighted \mcgbc\ problem.
\end{thm}

\paragraph*{LP-based algorithm}
This algorithm is based on solving the LP-relaxation of the IP formulation for \mec\ appearing in~\eqref{eq:ip-mec}. Once we have an optimal solution to the LP-relaxation, {\it we round the fractional solution into an integral solution}, with e.g., the probabilistic rounding technique of Srinivasan~\cite{srinivasan01distributions}.
We next sketch the basic idea of this probabilistic rounding technique.
Let $z^*$ be an optimal solution to the LP relaxation of~\eqref{eq:ip-mec}, and let $s$ be any sniffer. If $\sum_k z^*_{s,c} > 0$, one can view the induced solution $z^*_s:C \rightarrow [0,1]$ as a probability measure over the different channels (via normalization).
The goal is to decide on an integral channel assignment for $s$, namely, setting each $z^*_{s,c}$ to a value in $\set{0,1}$ such that {\em exactly} one variable out of the $k$ variables corresponding to sniffer $s$ is set to the value 1.
The algorithm builds a binary tree whose leaves corresponds to the $k$
variables $z_{s,k}$ associated with sniffer $s$, and pairs unset variables in a bottom-up fashion. The pairing is made such that an internal node sets at
least one of the variables corresponding to its children. This is done while
adjusting the (probability) value of the (other) unset variable.
This approach is proven to produce a valid assignment in linear time~\cite{srinivasan01distributions}. We refer to the above algorithm as \probrand.

\begin{thm}
\label{thm:probrand-approximation-ratio}
\probrand\ is a $(1-1/e)$-approximation algorithm for the weighted \mcgbc\ problem.
\end{thm}


We note that the approximation guarantee of the LP-based algorithms are best possible for the \mcgbc\ problem.
However, this lower bound does not necessarily hold for the \mec\ problem.

\section{QoM under the Sniffer-Centric Model}
\label{sec:sniffer-centric}
The user-centric model is more expressive than the sniffer-centric model, which assumes the availability of the binary observation matrix $\X$ only. However, we will show in this section the two models are intrinsically connected by devising algorithms to infer $\G$ and $\pp$ from $\X$.

Recall that in the sniffer-centric model, given an assignment $a \in \MA$,
$\prod_{k=1}^{K}{\MP_a(\xx_k)}$ is the probability distribution of {\bf binary}
observations
from $m$ sniffers. Let
$w(\xx_k)$ be the number of active users captured by sniffers in
channel $k$ given sniffer observations $\xx_k$.
The MEC problem under the sniffer-centric model is defined as follows.
\beq
\begin{array}{ll}
\underset{\pi(a)}{\max} & \sum_{a \in \MA} {\pi(a)\sum_{k=1}^{K}{\EE{w(\xx_k)}}} \\
\mbox{s.t.} & \pi(a) \in [0,1]  \\
& \sum_{a \in \MA}{\pi(a)} = 1,
\end{array}
\label{eq:sniffer_centric_opt}
\eeq
The expectation is with respective to $\MP_a(\xx_k)$.
QoM in sniffer-centric model can be explained as the expected number of active users
captured on all channels given the assignment probabilities.
Clearly, a pure strategy suffices, i.e., there exists an optimal assignment such that,
\beq
a^* = \argmax_a\sum_{k=1}^{K}{\EE{w(\xx_k)}}.
\label{eq:mec_sniffer}
\eeq

Even with pure strategies, the optimization problem defined in \eqref{eq:sniffer_centric_opt} is still challenging to solve directly. The main difficulty arises from the evaluation of $\EE{w(\xx_k)}$. Given $\xx_k$, one cannot decide how many users are active. Consider two scenarios. In the first case, two users are observed by two sniffers respectively. In the second case, a single user is observed by both sniffers. From binary observations alone, one cannot distinguish the two cases, which correspond to different number of active users. Furthermore, in contrast to the user-centric model, where transmission activities from different users are independent, observations of sniffers are correlated. As a result, $\MP_a(\xx_k)$ cannot be simplified as a product form.  This motivates us to exploit the underlying (though not directly observable) independence among users, and {\it map the optimization problem in sniffer-centric model to QoM under the user-centric model}.

In the sniffer-centric model, each sniffer only reports binary output
regarding the activities in the channel currently monitored by that sniffer, and thus the access probability of the users
as well as the bipartite graph $\GG$, are both {\em hidden}. Recall that
$\G$ refers to the adjacency binary matrix of $\GG$. We first
derive the sufficient and necessary conditions for unraveling the transmission
probabilities of the users given $\G$ and $\MP(\xx)$.

Let $\yy = \set{y_1,y_2,\ldots,y_n}^T$ be a vector of $n$ binary random variables,
where $y_j = 1$ if user $j$ transmits in its associated channel, and $y_j=0$ otherwise.  $\yy_{k}$
is the vector of activities for users transmitting on channel $k$ (i.e., users
in $U_k$).
The joint distribution of $\yy$ is given by $\MP(\yy) = \prod_{y_j=1} p_j \prod_{y_j=0} (1-p_j)$.
The product form is due to the independence among users' activities.
The main question we aim to answer is: given the vector $\xx_k$ of sniffers' observations, what knowledge can be obtained regarding
$\yy_{k}$?
Throughout this section, unless otherwise specified, we limit the
discussion to users and sniffers in a fixed channel $k$, and drop the
subscript. We will also denote by $g_{ij}$ the entry in the $i$'th row and $j$'th column of $\G$.

\begin{figure}[tp]
\begin{center}
\begin{tabular}{cc}
\includegraphics[width=1.6in]{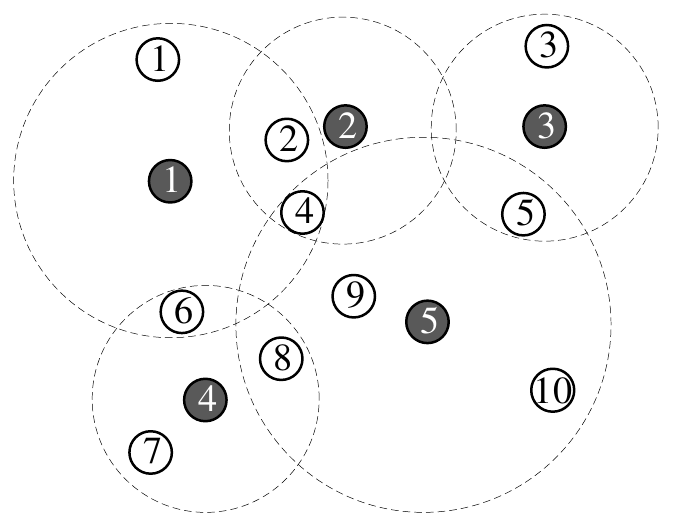} & \includegraphics[width=1.6in]{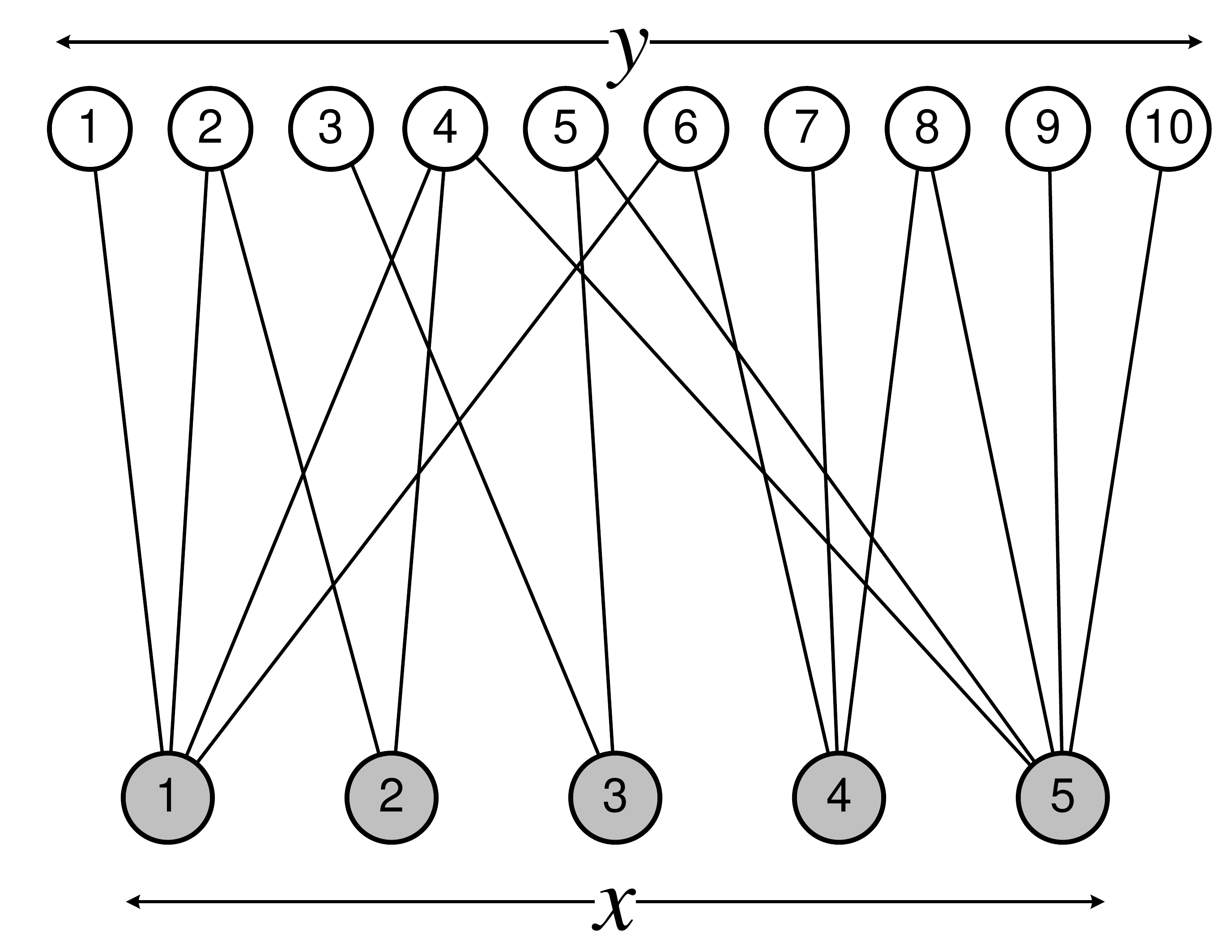}
\end{tabular}
$$
\G =
\begin{bmatrix}
1&\hspace{1.5mm}1&\hspace{1.5mm}0&\hspace{1.5mm}1&\hspace{1.5mm}0&\hspace{1.5mm}1&\hspace{1.5mm}0&\hspace{1.5mm}0&\hspace{1.5mm}0&\hspace{1.5mm}0\\
0&\hspace{1.5mm}1&\hspace{1.5mm}0&\hspace{1.5mm}1&\hspace{1.5mm}0&\hspace{1.5mm}0&\hspace{1.5mm}0&\hspace{1.5mm}0&\hspace{1.5mm}0&\hspace{1.5mm}0\\
0&\hspace{1.5mm}0&\hspace{1.5mm}1&\hspace{1.5mm}0&\hspace{1.5mm}1&\hspace{1.5mm}0&\hspace{1.5mm}0&\hspace{1.5mm}0&\hspace{1.5mm}0&\hspace{1.5mm}0\\
0&\hspace{1.5mm}0&\hspace{1.5mm}0&\hspace{1.5mm}0&\hspace{1.5mm}0&\hspace{1.5mm}1&\hspace{1.5mm}1&\hspace{1.5mm}1&\hspace{1.5mm}0&\hspace{1.5mm}0\\
0&\hspace{1.5mm}0&\hspace{1.5mm}0&\hspace{1.5mm}1&\hspace{1.5mm}1&\hspace{1.5mm}0&\hspace{1.5mm}0&\hspace{1.5mm}1&\hspace{1.5mm}1&\hspace{1.5mm}1
\end{bmatrix}
\begin{matrix}
x_1\\x_2\\x_3\\x_4\\x_5
\end{matrix}
\vspace{-2mm}
$$
$$
\hspace{5mm}
\begin{matrix}
y_1&y_2&y_3&y_4&y_5&y_6&y_7&y_8&y_9&y_{10}
\end{matrix}
$$
\end{center}
\caption{A sample network scenario with number of sniffers $m = 5$, number of users $n = 10$, its bipartite
graph transformation and its matrix representation. White circles represent independent users,
black circles represent sniffers and dashed lines illustrate sniffers' coverage range.}
\label{fig:mt}
\end{figure}

Using the adjacency matrix, and using $\wedge$ to represent Boolean {\em AND} and $\vee$ to represent Boolean {\em OR}, we have the following:
\beq
x_i = \bigvee_{j=1}^{n}{g_{ij}\wedge y_j}, \mbox{ $i = 1, \ldots, m$},
\label{eq:boolean}
\eeq
i.e., $x_i=1$ iff there exists a user $j$ within the range of sniffer $i$ ($g_{ij}=1$) that transmits ($y_j=1$).
Define the set $\Y(\xx) = \{\yy\mid \bigvee_{j=1}^{n}{g_{ij}\wedge y_j} = x_i, \forall i\},$
i.e., the set of user activity profiles that are consistent with the sniffers' observations.
Therefore,
\beq
\MP(\xx) = \MP(\yy \in \Y(\xx)) = \sum_{\yy \in \Y(\xx)} \MP(\yy)
\label{eq:relation}
\eeq
An example network with sniffers and users, the corresponding bipartite graph, and its matrix representation $\G$ are given in Figure~\ref{fig:mt}.

\subsection{Relationship between the user-centric and sniffer-centric models with known $\G$ and unknown $\pp$}
\label{sec:sniffer-centric-vs-user-centric}
The necessary and sufficient conditions that uniquely determine $\pp$ using
$\G$ and $\MP(\xx)$ is characterized in the following theorem.
\begin{thm}
\label{thm:distribution-characterization}
Given $\GG=(S,U,E)$, $\pp$ can be uniquely determined by
$\MP(\xx)$ {\em iff} $\forall u_j\neq u_{j'} \in U$, $N(u_j) \neq N(u_{j'})$.
\end{thm}
\begin{IEEEproof}
\begin{itemize}
\item[$\Rightarrow$] It is easy to see that the necessary condition holds. If two
users have the same set of sniffer neighbors, unless packet headers are
analyzed, their activities cannot be distinguished.

\item[$\Leftarrow$] To prove the sufficient condition, we construct a procedure
to determine $\pp$ from sniffer's joint distribution $\MP(\xx)$.

\noindent \underline{Case 1:} First, we consider a more restrictive constraint,
namely, $\forall u_j \neq u_{j'} \in U$, $N(u_j) \not\subseteq N(u_{j'})$.  We let
$g_j$ denote the $j$'th column of the adjacency matrix $\G$, i.e., a binary vector of
length $m$.
In other words, $g_j$ is the coverage vector of the $j$'th sniffer.
Since $\forall u_j \neq u_{j'} \in U$, $N(u_j) \not\subseteq N(u_{j'})$,
we have $g_j \cup g_{j'} \neq g_{j}$ and $g_{j} \cup g_{j'} \neq g_{j'}$.
From~\eqref{eq:relation}, we have
$$\MP(\xx =g_j) = p_j \prod_{j' \in U, j'\neq j}{(1-p_{j'})},$$
Since $\MP(\xx =\emptyset) = \prod_{j \in U}{(1-p_j)}$ (recall by our abuse of notation that $\emptyset$ is
the all-zero vector of length $m$), we have
\beq
p_j = \frac{\MP(\xx =g_j)}{\MP(\xx
= g_j) + \MP(\xx =\emptyset)}.
\label{eq:pj_calc}
\eeq

\noindent \underline{Case 2:} Now we consider the case when the condition
$\forall u_j \neq u_{j'} \in U$, $N(u_j) \not\subseteq N(u_{j'})$ is violated.
Without loss of generality, assume only one such pair $(j,j')$ exist and $N(u_{j'})
\subseteq N(u_j)$ (analysis for more complicated cases follows the same
line of arguments). In this case, $p_{j'}$ can be derived as in the previous case.
However, equation~(\ref{eq:pj_calc}) does not hold for user $j$ any more since if $\xx
=g_j$, user $j'$ may or may not be active. More specifically,
\beq
\MP(\xx=g_j) = p_j\prod_{j'' \in U, j''\neq j,j'}{(1-p_{j''})}.
\eeq
Therefore, we have
\begin{equation}
p_j = \frac{\MP(\xx =g_j)(1-p_{j'})}{\MP(\xx
=g_j)(1-p_{j'}) + \MP(\xx =\emptyset)}.
\label{eq:recur}
\end{equation}

In other words, the active probability of users can be computed by considering
the users with the smallest degree in $\G$ first, and then applying~\eqref{eq:recur}
iteratively in ascending order of node degree.
\end{itemize}
\end{IEEEproof}

The above theorem essentially shows that in the sniffer-centric model, if
$\G$ is known, then one can effectively determine the transmission
probabilities of the users. In presence of measurement noise, methods such as
Expectation-Maximization can be applied.  We therefore obtain an instance of the
problem corresponding to our user-centric model, which can be solved
efficiently using the algorithms described in Section~\ref{sec:mec-algorithms}.
%

	
\paragraph*{Comment} Though Theorem~\ref{thm:distribution-characterization} requires the users be connected to different sets of sniffers, violation of the condition would not affect the channel assignment. For example, when two users $u$ and $v$ are connected to the same set of sniffers, and are thus ``indistinguishable" in the binary sniffer observations, we can effectively view them as a single user with active probability $1 - (1-p_u)(1-p_v)$ if users are active independently, or $p_u + p_v$ if only one user can be active at a time (e.g., due to CSMA).

\subsection{Inference of unknown $\G$ and $\pp$ using binary ICA}
\label{sec:sniffer-centric-unknown-graph}
\begin{figure*}[th]
\begin{center}
\begin{tabular}{c}
\includegraphics[width=6.5in]{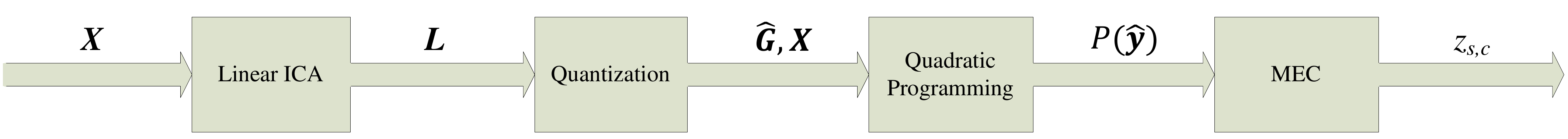} \\
(a) Quantized Linear ICA \\\\
\includegraphics[width=6.5in]{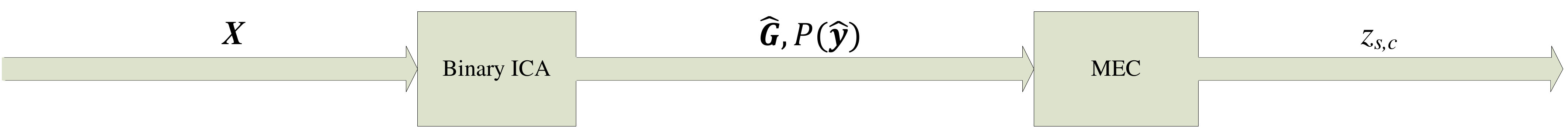} \\
(b) Binary ICA
\end{tabular}
\caption{Channel selection algorithm under \qlica\ and \bica\ models}
\label{fig:final-algorithm}
\end{center}
\end{figure*}
%
In this section, we derive methods to estimate the unknown mixing matrix $\G$ and
the active probability vector $\pp$.
Consider again the example in Figure~\ref{fig:toy}. Let sniffers $s_1$ and $s_2$ be assigned to channel 1 and
observe the activity of a single user $y_1$. In this case, $x_1 = x_2$. Therefore,
$$
\begin{array}{lll}
\MP(\xx) & = & \MP(x_1)\one{x_1 = x_2}\\ & = & \MP(y_1 = x_1) \\ & = & \left\{\begin{array}{ll} p_1, & x_1 = 1 \\ 1-p_1, & x_1 = 0 \end{array}\right..
\end{array}
$$
Therefore, if the joint distribution of $\xx_k$ is the product of a marginal
distribution with an indicator function, and the two marginal distributions are
identical, we can infer that both sniffers observe the same set of users.
Generally, the joint distribution of $\xx$ preserves a certain stochastic
``structure'' of the user's activities. We will formalize this observation in the
subsequent section by devising two inference methods to estimate $\G$ and $\pp$ from $\MP(\xx)$.

\subsubsection{Quantized Linear ICA (\qlica)}
First we will estimate $\G$ by applying the classic ICA on the binary data
followed by a quantization process. Then $\MP{(\yy)}$ can then be calculated by
solving a quadratic programming problem.
\paragraph*{Estimation of $\G$}
The problem is similar to what was addressed by the Independent Component Analysis (ICA) scheme~\cite{Hyvarinen00},
where the observed data is expressed as a linear transformation of latent
variables that are non-Gaussian and mutually independent.
Classic ICA assumes that both $\yy$ and $\xx$ are continuous random variables and that $\xx$ is the outcome of a linear mixing of $\yy$, and thus is not
directly applicable to our problem.
We adopt the algorithm presented in
\cite{himberg01} with some modifications. The basic idea
is as follows.

We first observe that~\eqref{eq:boolean} can be simplified using linear mixing and a (coordinate-wise) unit step
function.
\beq
\xx = \unitstep(\G\yy),
\eeq
where $\unitstep(\cdot)$ is a unit step function defined by
$\unitstep(r) = \one{r > 0}$.
By applying the standard ICA on $\xx$, we can ``decompose'' the observation to $\xx \approx \L \sss$, with $\L$ is the {\em linear} mixing matrix and $\sss$ is the collection of random sources. However, both $\L$ and $\sss$ are not the solutions to our problem since
they contain fractional values.
Therefore, we {\it quantize} $\L$ to get the inferred {\it binary} mixing matrix $\hat{\G}$
as follow,
\beq
\hat{\G}=\unitstep(\L \Lambda^{-1}-\T).
\label{eq:quantization}
\eeq
$\Lambda$ is the diagonal scaling matrix with $\lambda_{ii} = \maxstep(\ell_i)$,
where $\ell_i$ is the $i$'th column of $\L$, and
\beq
\maxstep(\rr) = \left\{\begin{array}{ll} \max(\rr) & \quad \mbox{if } |\max(\rr)|> |\min(\rr)| \\
\min(\rr) & \quad \mbox{otherwise.} \end{array}\right.
\eeq
$\Lambda$ scales the elements in the mixing matrix to the maximum value 1. The
matrix $\T$ contains thresholds, such that the higher the threshold value, the sparser $\hat{\G}$ is.


%

\paragraph*{Estimation of $\MP{(\yy)}$}
Once $\hat{\G}$ is determined, $\MP{(\yy)}$ needs to be estimated. From $x_i =
U(\hat{g}_i y_i)$, where $\hat{g}_i$ is the $i$'th row of $\hat{\G}$ (i.e., the estimated coverage vector of sniffer $s_i$), we have,
\beq
p(x_i = 0) = \prod_{\hat{g}_{ij} = 1}{p(y_j = 0)}.
\eeq
The product is due to the
independence of $y_i$'s. Taking $\log(\cdot)$ on both sides, we have
\beq
\log(p(x_i = 0)) = \sum_{\hat{g}_{ij} = 1}{\log(p(y_j = 0))}.
\eeq
Let $\alpha_i
= \log(p(x_i = 0))$, and $\beta_i = \log(p(y_j = 0))$. Define $\mathbf{\alpha}
= \{\alpha_1, \alpha_2, \ldots, \alpha_m\}^T$, and $\mathbf{\beta} = \{\beta_1,
\beta_2, \ldots, \beta_n\}^T$. We can calculate $p(y_j=0)$ (and consequently obtain $\MP{(\yy)}$) by solving the following optimization
problem.
\beq
\begin{array}{cc}
\underset{\beta}{\min} & \parallel\mathbf{\alpha} - \hat{\G}\mathbf{\beta}\parallel^2 \\
\mbox{s.t.} & \mathbf{\beta} < 0,
\end{array}
\label{eq:optimization}
\eeq
where $\parallel\cdot\parallel$ is the second norm of a vector.
The objective function minimizes the distance between the real observation vector $\xx$ and its reconstructed counterpart ($\hat{\G}\beta$).
Clearly, this is a constrained quadratic programming problem with a positive
semi-definite matrix (i.e., all eigenvalues are non-negative), and can be
solved in polynomial time.

\paragraph*{Channel selection}
With the estimated $\hat{\pp}$ and $\hat{\G}$ at hand, we effectively transform the sniffer-centric
model to the user-centric model. Methods described in Section~\ref{sec:mec-algorithms} can
then be applied to determine the channel assignment of each sniffer.
\verb QuantizeICA ~algorithm which infers $\hat{\pp}$ and $\hat{\G}$ is presented in Algorithm~\ref{algo:qlica}
and the complete \qlica\ scheme is illustrated in Figure~\ref{fig:final-algorithm}(a).
\begin{algorithm}[h]
\caption{Quantized linear ICA inference}
\footnotesize
\label{algo:qlica}
\SetKwData{Left}{left}
\SetKwInOut{Input}{input}
\SetKwInOut{Output}{output}
\SetKwInOut{Init}{init}
\SetKwFor{For}{for}{do}{endfor}
\SetKwFunction{QuantizeICA}{QuantizeICA}
\QuantizeICA($\X$)\\
\Input{Data matrix $\X_{m\times T}$}
\Init{$\T = $ threshold matrix\;}
\nl $\L = $ mixing matrix obtained by applying ICA on $\X$\;
\nl $\Lambda = $ diagonal scaling matrix calculated from $\L$\;
\nl $\hat{\G}=\unitstep(\L \Lambda^{-1}-\T)$\;
\nl Calculate $\alpha$ from $\X$ with $\alpha_i = \log(p(x_i = 0))$\;
\nl Obtain $\hat{\pp}$ by solving the quadratic programming problem in (\ref{eq:optimization})\;
\nl\Output{$\hat{\pp}$ and $\hat{\G}$}
\end{algorithm}
\normalsize
\paragraph*{A toy example}
We next give a simple example, which provides insight as
to the operations of \qlica. Let us reconsider the network in Figure~\ref{fig:toy} with $u_1$ and $u_2$ operate on one single channel.
With $T = 10$ observations, supposedly we have the activity matrix

\begin{center}
\vspace{-0.2in}
$$
\Y =
\begin{bmatrix}
0&0&1&0&0&0&0&0&1&0\\
0&1&1&0&0&1&1&1&0&0
\end{bmatrix}.
$$
\end{center}

$y_{ij} = 1$ indicates that user $y_i$ is active on the channel at time slot $j$. $\Y$
is hidden and unknown to us. Since $\G = \begin{bmatrix} 1&0\\ 1&1 \end{bmatrix}$,
we have the observation matrix

\begin{center}
\vspace{-0.2in}
$$
\X =
\begin{bmatrix}
0&0&1&0&0&0&0&0&1&0\\
0&1&1&0&0&1&1&1&1&0
\end{bmatrix}.
$$
\end{center}

Applying the linear ICA, we obtain

\begin{center}
\vspace{-0.2in}
$$
\L =
\begin{bmatrix}
0.30&-0.35 \\
-0.20&-0.35 \\
\end{bmatrix},
\Lambda^{-1} =
\begin{bmatrix}
-2.89&0\\
0&-2.89\\
\end{bmatrix}.
$$
\end{center}

With threshold $\T = 0.5$, solving the equation (\ref{eq:quantization}) and the optimization
problem (\ref{eq:optimization}), we have the following inferred results.
\begin{center}
\vspace{-0.2in}
$$
\hat{\G} =
\begin{bmatrix}
0&1\\
1&1
\end{bmatrix},
\hat{\pp} = \{0.5, 0.2\}.
$$
\end{center}

Inferred results $\hat{\G}$ and $\hat{\pp}$ are actually permutations of the original
mixing matrix $\G$ and the active probability $\pp$. We see that \qlica\ can successfully infer
information regarding the underlying model from $\MP(\xx)$.
%
%

\vspace{0.5cm}

\subsubsection{Binary ICA (\bica)}
Instead of applying a quantization process on the result of linear ICA, we can
apply the \bica~algorithm proposed in \cite{Nguyen2011TSP} to determine
$\MP(\yy)$ and $\G$ by exploiting the {\em OR} mixture model between $\yy$ and
the observation variable $\xx$. Compared with \qlica, \bica~explicitly account
for the generative model and thus leads to more accurate estimation results.
However, this comes at the expense of higher computation complexity. In the
worst case, given $m$ sniffers, the run time of the algorithm is
$O(m2^m)$.\footnote{Several techniques are suggested in \cite{Nguyen2011TSP} to
reduce the computation complexity.} For completeness, we first define some
notation and then outline the \bica~algorithms.

\paragraph*{Joint estimation of $\G$ and $\MP(\yy)$}
The basic idea of \bica~algorithm is as follow: given
an observation matrix $\X$ from $m$ sniffers, we will first assume that there exists
at most $2^m$ distinguishable users. Each user is represented in $\G$ by a unique
column $\in \{0,1\}^m$ indicating its connections to $m$ sniffers. $2^m$ users
are ordered by ascending values of their corresponding columns. We will recursively construct 2 submatrices from $\X$
such that the first submatrix captures the joint distribution of activities from
the first $2^{m-1}$ users (in product form due to the independence assumption).
If the submatrix is small enough, the join distribution can be inferred
directly, otherwise, a divide-and-conquer approach is taken. Once the join distribution of the first $2^{m-1}$
users are available, that of the remaining $2^{m-1}$ users can be inferred
from the second submatrix.

Next, we present in detail the proposed \bica~algorithm.
Let define $\X_{(h-1)\times T}^0$ to be a submatrix of $\X$,
where the rows correspond to observations of $x_1, x_2, \ldots, x_{h-1}$
for $t = 1, 2, \ldots, T$ such that $x_{ht} = 0$, i.e., the first submatrix
mentioned above. Also, define $\X_{(h-1)\times T}$ to
be the matrix consisting the first $h-1$ rows of $\X$, i.e., the second
submatrix. Let $\MF(.)$ be the frequency function of some event,
we have the iterative inference algorithm as illustrated in Algorithm~\ref{algo:inc}.

\begin{algorithm}[h]
\caption{Incremental binary ICA inference}
\footnotesize
\label{algo:inc}
\SetKwData{Left}{left}
\SetKwInOut{Input}{input}
\SetKwInOut{Output}{output}
\SetKwInOut{Init}{init}
\SetKwFor{For}{for}{do}{endfor}
\SetKwFunction{FindBICA}{FindBICA}
\FindBICA($\X$)\\
\Input{Data matrix $\X_{m\times T}$}
\Init{$n = 2^{m} -1$\;
$\hat{\pp}$ = $1\times n$ zero vector\;
$\hat{\G}$ = $m \times (2^{m} -1)$ matrix with rows corresponding all possible binary vectors of length $m$\;
$\varepsilon$ = the minimum threshold for $p_h$ to be considered a real component;} \BlankLine

\nl\eIf{$m = 1$} {\label{line1}
\nl    $\hat{p_1} = \MF(x_1 = 0)$\;\label{line2}
\nl    $\hat{p_2} = \MF(x_1 = 1)$\;\label{line3}
    } {
\nl    \eIf{$\X^0_{(m-1)\times T} = \emptyset$} {\label{line12}
\nl        $\hat{p}_{1 \ldots 2^{m-1}}$ = \FindBICA($\X_{(m-1)\times T}$)\;\label{line13}
\nl        $\hat{p}_{2^{m-1}+1} = 1$\;\label{line14}
\nl        $\hat{p}_{2^{m-1}+2\ldots 2^m} = 0$\;\label{line15}
        } {
\nl    $\hat{p}_{1 \ldots 2^{m-1}}$ = \FindBICA($\X^0_{(m-1)\times T}$)\;\label{line4}
\nl    $\hat{p}'_{1 \ldots 2^{m-1}}$ = \FindBICA($\X_{(m-1)\times T}$)\;\label{line5}
\nl    \For{$l = 2, \ldots, 2^{m-1}$}{\label{line6}
\nl        $\hat{p}_{l+2^{m-1}} = 1-\frac{1-\hat{p}'_l}{1-\hat{p}_l}$\;\label{line7}
       }
\nl    $\hat{p}_{2^{m-1}+1} = \frac{\MF{(x_m=1 \wedge x_i=0, \forall i \in [1 \ldots m-1])}}{\prod_{l=1 \ldots 2^m-1, l \neq 2^{m-1}+1}{(1-\hat{p}_l)}}$\;\label{line8}
}
}
\nl\For{$h = 1, \ldots, 2^{m}$}{\label{line9}
\nl    \If{$(\hat{p}_h < \varepsilon) \vee (\hat{p}_h = 0)$}{\label{line10}
\nl            prune $\hat{p}_h$ and corresponding column $\hat{g}_{h}$\;\label{line11}
        }
}
\nl\Output{$\hat{\pp}$ and $\hat{\G}$}
\end{algorithm}
\normalsize

When the number of observation variables $m=1$, there are only two possible unique sources,
one that can be detected by the monitor $x_1$, denoted by [1]; and one that cannot,
denoted by [0]. Their active probabilities can easily be calculated by counting
the frequency of $(x_1 = 1)$ and $(x_1 = 0)$ (lines~\ref{line1} --
\ref{line3}). If $m \ge 2$, $\pp$ and $\G$ are estimated through a recursive process.
$\X^0_{(m-1)\times T}$ is sampled from columns of $\X$ that have $x_m = 0$. If $\X^0_{(m-1)\times T}$
is an empty set (which means $x_{mt} = 1, \forall t$) then we can associate $x_m$
with a constantly active component and set the other components' probability accordingly
(lines~\ref{line12} -- \ref{line15}). If $\X^0_{(m-1)\times T}$ is non-empty, we invoke \verb FindBICA ~on
two sub-matrices $\X^0_{(m-1)\times T}$ and $\X_{(m-1)\times T}$ to determine
$\hat{p}_{1 \ldots 2^{m-1}}$ and $\hat{p}'_{1 \ldots 2^{m-1}}$, then
infer $\hat{p}_{2^{m-1}+1 \ldots 2^{m}}$ (lines~\ref{line4} --
\ref{line8}). Finally, $\hat{p}_h$ and its corresponding column $\hat{g}_h$ in $\hat{\G}$ are
pruned in the final result if $\hat{p}_h < \varepsilon$ (lines~\ref{line9} --
\ref{line11}).

\paragraph*{Channel selection}
Now that $\hat{\pp}$ and $\hat{\G}$ are inferred, we again transform the sniffer-centric
model to the user-centric model and apply methods Section~\ref{sec:mec-algorithms} to find
optimal sniffer-channel assignment scenario.
The complete \bica\ scheme is illustrated in Figure~\ref{fig:final-algorithm}(b).
\paragraph*{A toy example}
Again, let us reconsider the network in Figure~\ref{fig:toy}.
Recall that $\pp = \{0.2, 0.5\}$, $T = 10$ and the observation matrix is

\begin{center}
\vspace{-0.2in}
$$
\X =
\begin{bmatrix}
0&0&1&0&0&0&0&0&1&0\\
0&1&1&0&0&1&1&1&1&0
\end{bmatrix}.
$$
\end{center}

Following the \verb FindBICA ~algorithm,
we will first decompose $\X$ into $\X^0_{1\times T}$ and $\X_{1\times T}$.

\begin{center}
\vspace{-0.2in}
$$
\X^0_{1\times T} =
\begin{bmatrix}
0&0&0&0
\end{bmatrix},
$$
$$
\X_{1\times T} =
\begin{bmatrix}
0&0&1&0&0&0&0&0&1&0
\end{bmatrix}.
$$
\end{center}

From $\X^0_{1\times T}$, we have $\hat{p}_0 = 1$ and $\hat{p}_1 = 0$. Similarly from
$\X_{1\times T}$, we have $\hat{p}'_0 = 0.8$ and $\hat{p}'_1 = 0.2$. Applying the procedure
in Algorithm \ref{algo:inc}, we have the inferred results:

\begin{center}
\vspace{-0.2in}
$$
\hat{\G} =
\begin{bmatrix}
0&1&0&1\\
0&0&1&1
\end{bmatrix},
\hat{\pp} = \{1, 0, 0.5, 0.2\}.
$$
\end{center}

The first and second column of $\hat{\G}$ and $\hat{\pp}$ will be pruned
since we are not interested in components that cannot be observed (with $\hat{g}_i = 0$)
or components with their active probabilities smaller than
$\varepsilon$ ($\varepsilon$ is set to be 0.01 in this case). Thus yields the final result to be exactly equal to
the ground truth. Toy example shows that both \qlica\ and \bica\
can accurately predict the underlying model without priory
knowledge on the latent variables' activities in simple cases. The next section
will provide an extensive evaluation on performance of \qlica\ and \bica.
\section{Simulation Validation}
\label{sec:evaluation}

In this section we evaluate the performance of different algorithms under the
user-centric and sniffer-centric models using both synthetic and real traces.
Synthetic traces allow us to control the parameter settings while real network traces provide insights on the performance under realistic traffic loads and user distributions.

In addition to the \greedy\ and LP-based algorithm, we also consider {\bf
Max Sniffer Channel (Max)} where a sniffer is assigned to its busiest channel.
This scheme is the most intuitive approach in practical networks where the user model
is not available and sniffers have to decide their channel assignment {\it
non-cooperatively} based on local observations. Note it is easy to construct
scenarios where Max performs arbitrarily bad. Thus, its worst case performance
is unbounded.
For the inference scheme in the \qlica\ model, we used the FastICA
algorithm~\cite{Hyvarinen00} to compute the linear mixing matrix $\L$.

\subsection{QoM under different models}

\subsubsection{Synthetic traces}

\begin{figure}[htb]
\begin{center}
\includegraphics[width=3.0in]{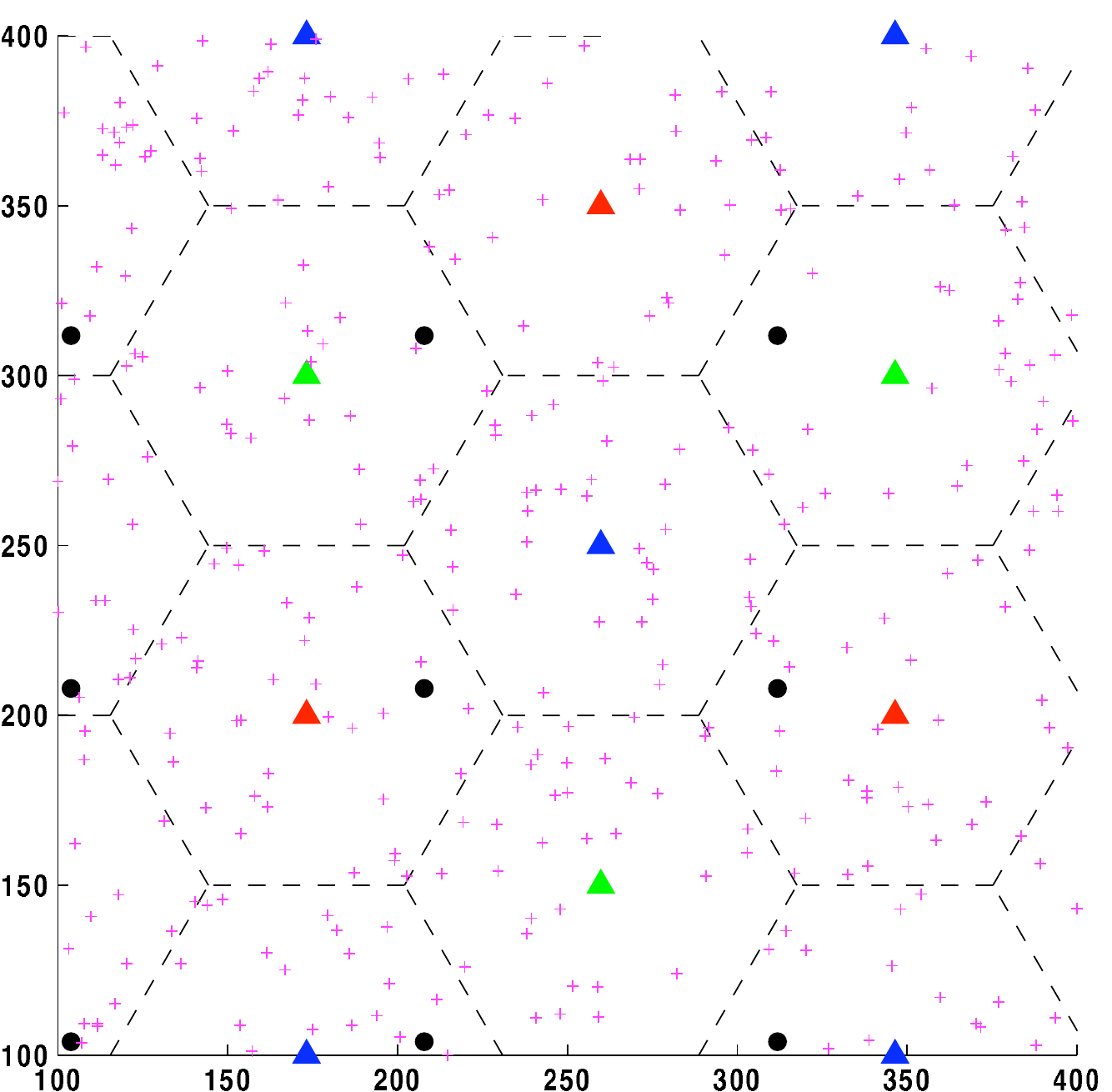}
\caption{Hexagonal layout with users (`+'), sniffers (solid dots), base stations (triangles), and channels of each cell (in different triangle colors)}
\label{fig:deployment}
\end{center}
\end{figure}

\begin{figure*}[t!]
\begin{center}
\begin{tabular}{ccc}
\includegraphics[width=2.2in]{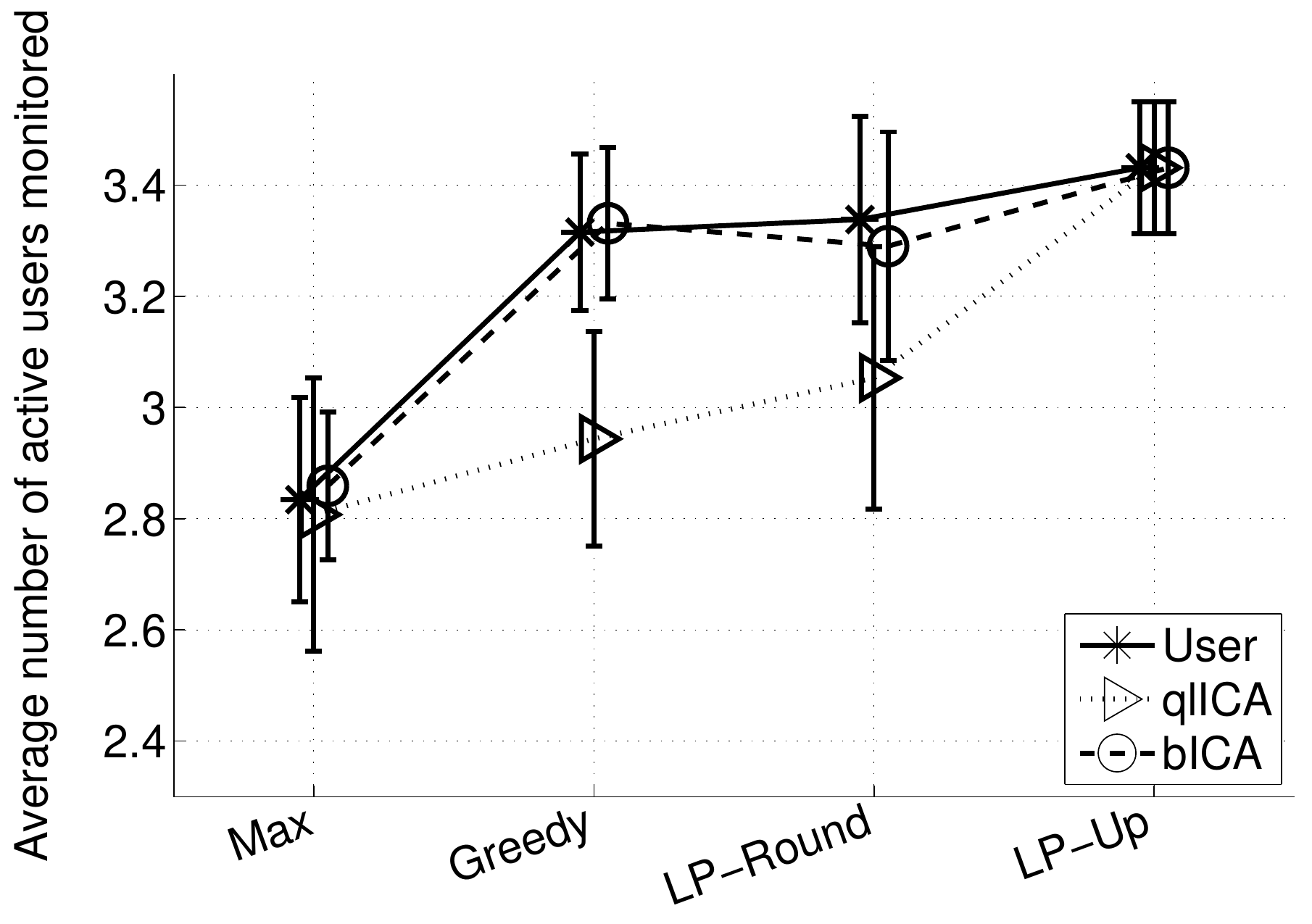} &
\includegraphics[width=2.2in]{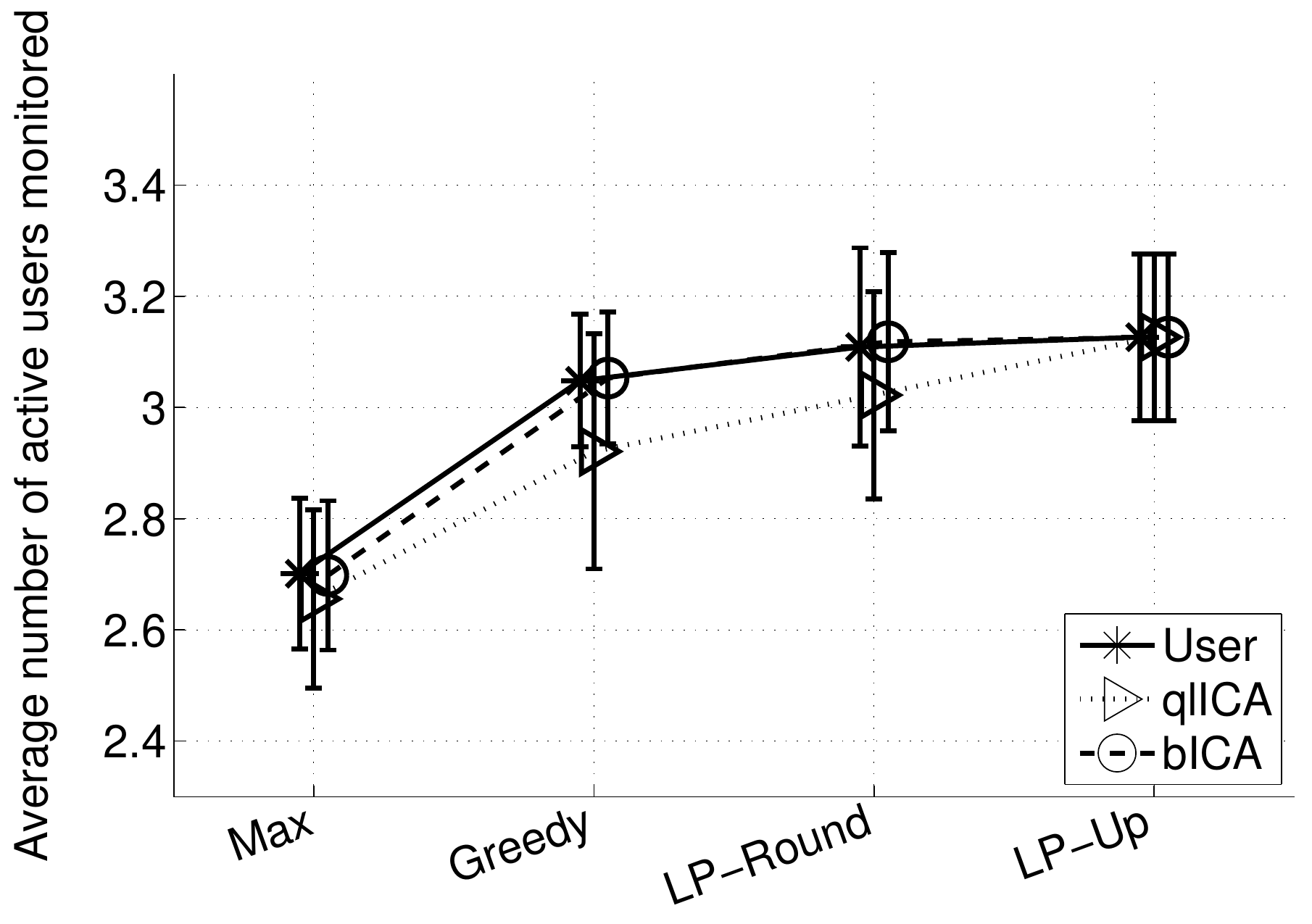} &
\includegraphics[width=2.2in]{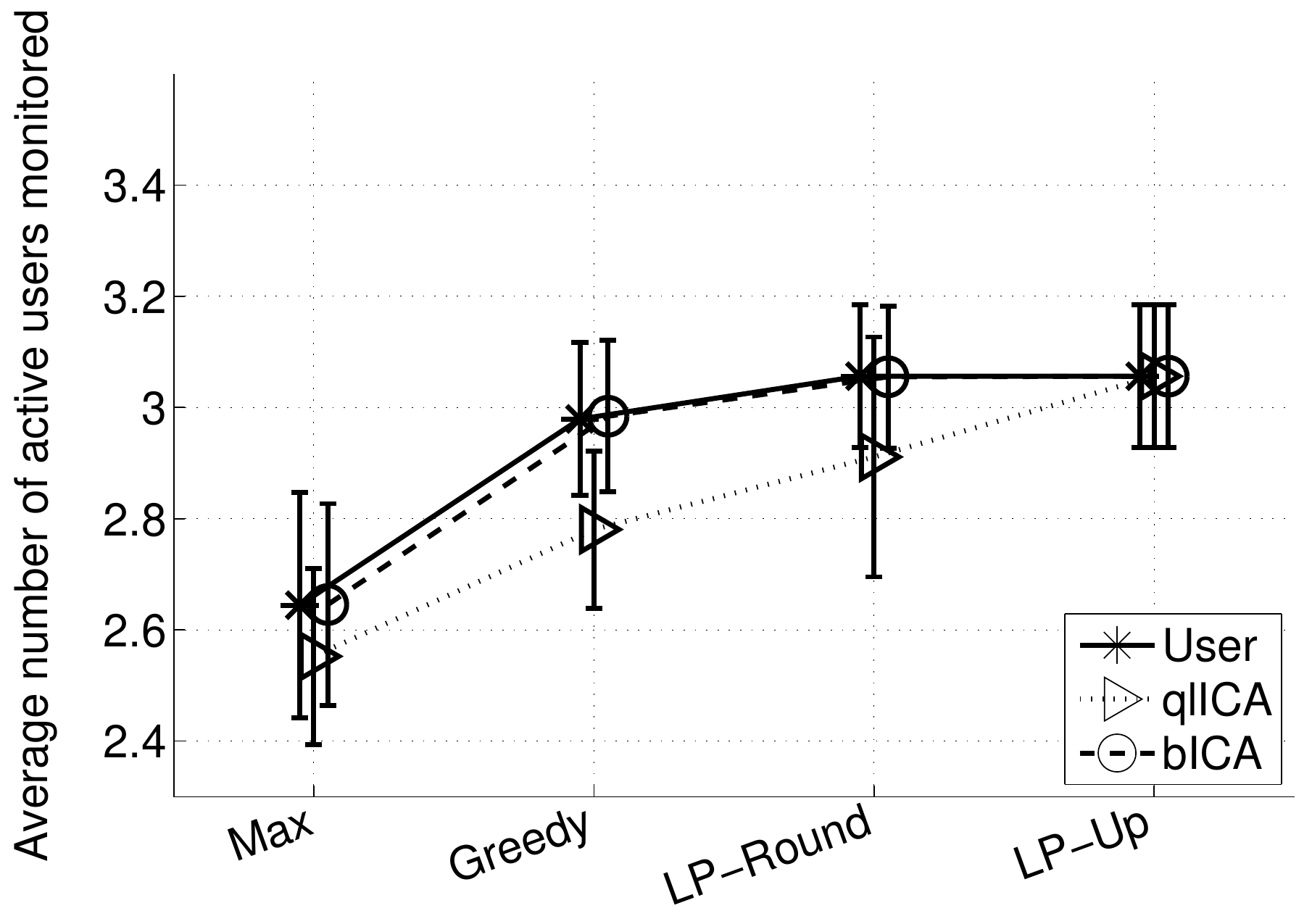} \\
(a) 3 channels &
(b) 6 channels &
(c) 9 channels
\end{tabular}
\caption{QoM under three models: the user-centric model (User), \qlica\ and \bica\ with 3, 6 and 9-channel synthetic traces}
\label{fig:synthetic_3_6_9_120}
\end{center}
\end{figure*}

In this set of simulations, 500 wireless users are placed randomly in a
$500\times 500$ square meter area. The area is partitioned into hexagon cells with
circumcircle of radius 86 meters. Each cell is associated with a base station
operating in a channel (and so are the users in the cell). The channel to
base station assignment ensures that {\it no neighboring cells use the same channel}.
25 Sniffers are deployed in a grid formation separated by distance 100 meters,
with a coverage radius of $120$ meters.
A snap shot of the synthetic deployment is
shown in Figure~\ref{fig:deployment}.  The transmission probability of users is selected
uniformly in (0, 0.06], resulting in an average busy probability of 0.2685 in
each cell. Threshold {\bm $T$} for \qlica\ is set at 0.5 and threshold $\varepsilon$
for \bica\ is set at 0.01.
We vary the total number of orthogonal channels from 3 to
9.\footnote{In 802.11a networks, there are 8 orthogonal channels in 5.18-5.4GHz, and one in 5.75GHz.} The results shown are the average of 20 runs with different seeds.

Figure~\ref{fig:synthetic_3_6_9_120} shows QoM calculated by three
algorithms (Max, Greedy and LP-Round) and the theoretical upper bound (LP-Up)
on two models using synthetic traces of 3, 6, 9 channels, respectively.
Results of the user-centric model are shown in solid lines while results of different inference algorithms
(e.g., \qlica\ and \bica) in the sniffer-centric model are shown in dotted and dashed lines, respectively.
In the user-centric model, one can see that the
performance of Greedy and the LP-based algorithm with random rounding are
comparable to LP-Up, and both outperform Max in all
three traces. Recall that according to Max, a sniffer
non-cooperatively decides its own channel assignment and selects the most
active channel. Clearly, Max does not take into account the
correlations among the observations of neighboring sniffers in the same
channel. In contrast, in the sniffer centric case, the proposed inference
algorithms can indeed extract such a correlative structure from the binary
observations as shown by their superior performance over Max.

Additionally, we observe that the expected number of users monitored by the
algorithms using \bica\ is higher than that of \qlica\ and is very
close to that attained in the user centric model (where we assumed
to have complete knowledge of users' activities and their relationship to sniffers).
This indicates that \bica\
algorithm indeed produces inferred models that are very close to the ground truths.
Having a good estimation of $\hat{\G}$\footnote{A predicted user in $\hat{\G}$ is
actually the aggregation of real users in a unique sniffer coverage area since
we simply cannot distinguish between different users that can only be monitored by the
same set of sniffers.} and $\hat{\pp}$ as the input, Greedy and LP-Round can produce
channel assignments whose performance is close to LP-Up.

We further note that by comparing results from
Figure~\ref{fig:synthetic_3_6_9_120}(a) to
Figure~\ref{fig:synthetic_3_6_9_120}(c), the QoM metric reduces as the total
number of channels increases for all schemes, including LP-Up.
This is due to the fact that users scatter over more channels,
and a fixed number of sniffers is no longer sufficient to provide good coverage.

\subsubsection{Real traces}

In this section, we evaluate our proposed schemes using real traces collected
from the UH campus wireless network using 21 WiFi sniffers deployed in the
Philip G. Hall. Over a period of 6 hours, between 12 p.m. and 6 p.m., each sniffer
captured approximately 300,000 MAC frames.
Altogether, 655 unique users are observed operating over three
channels.\footnote{Our measurements used the campus IEEE 802.11g WLAN, which
has three orthogonal channels.} The number of users observed on WiFi channels 1, 6,
11 are 382, 118, and 155, respectively.
The histogram of user active
probability (calculated as the percentage of 20$\mu$s slots that a user is
active) is shown in Figure~\ref{fig:hist}.  Clearly, most users are active less
than 1\% of the time except for a few heavy hitters.
The average user active probability is 0.0014.

\begin{figure*}[t!]
\begin{center}
\begin{tabular}{ccc}
\includegraphics[width=2.2in]{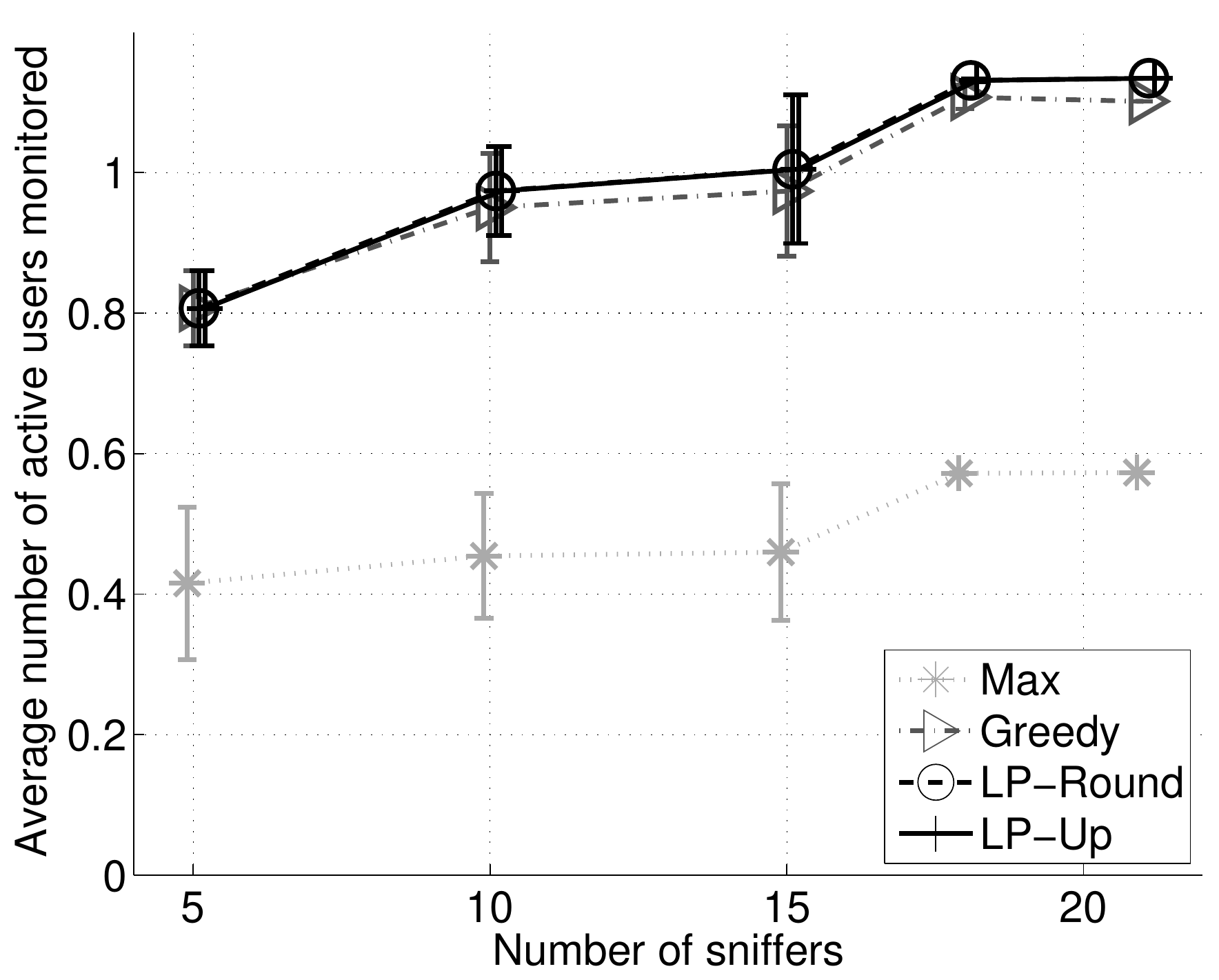} &
\includegraphics[width=2.2in]{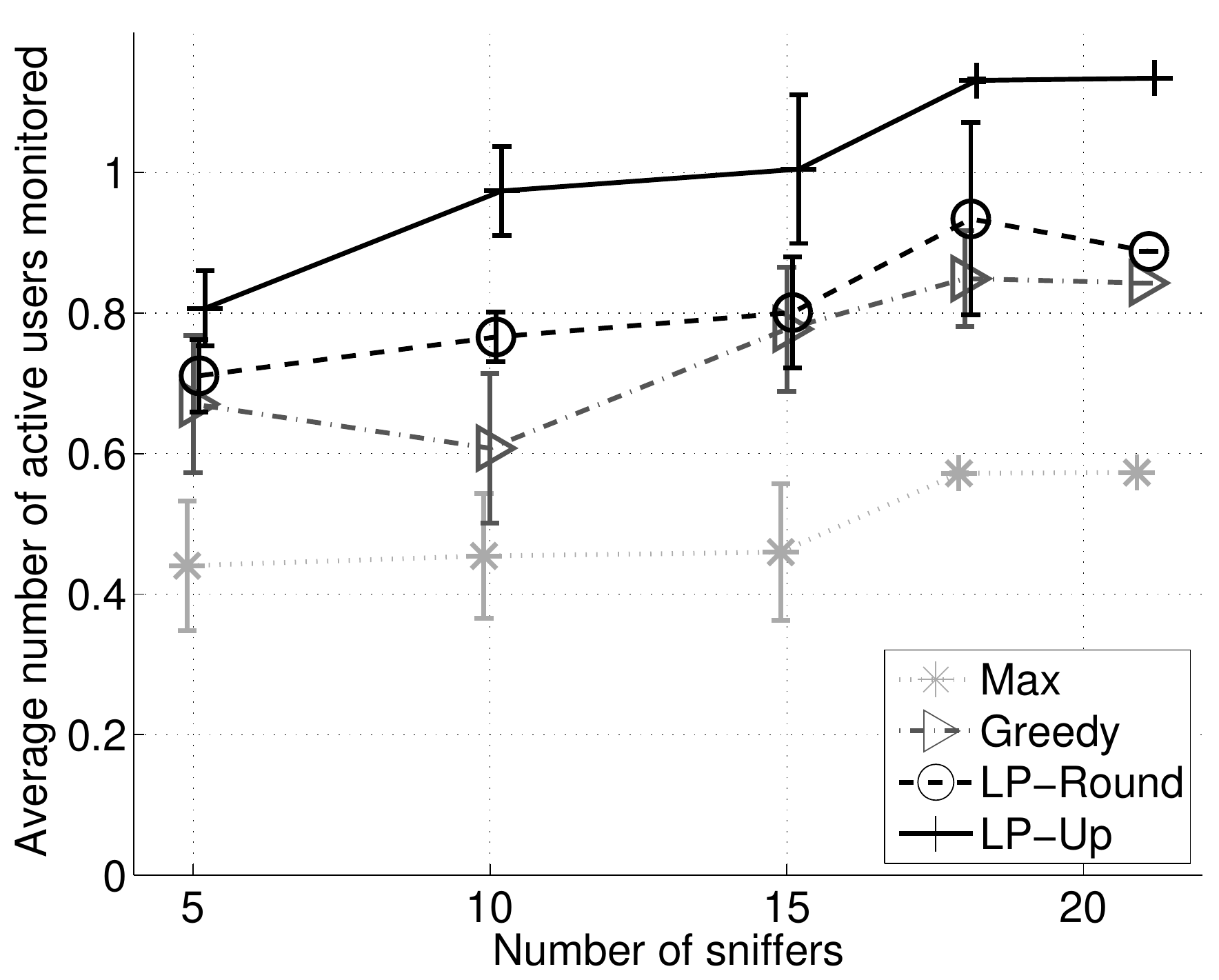} &
\includegraphics[width=2.2in]{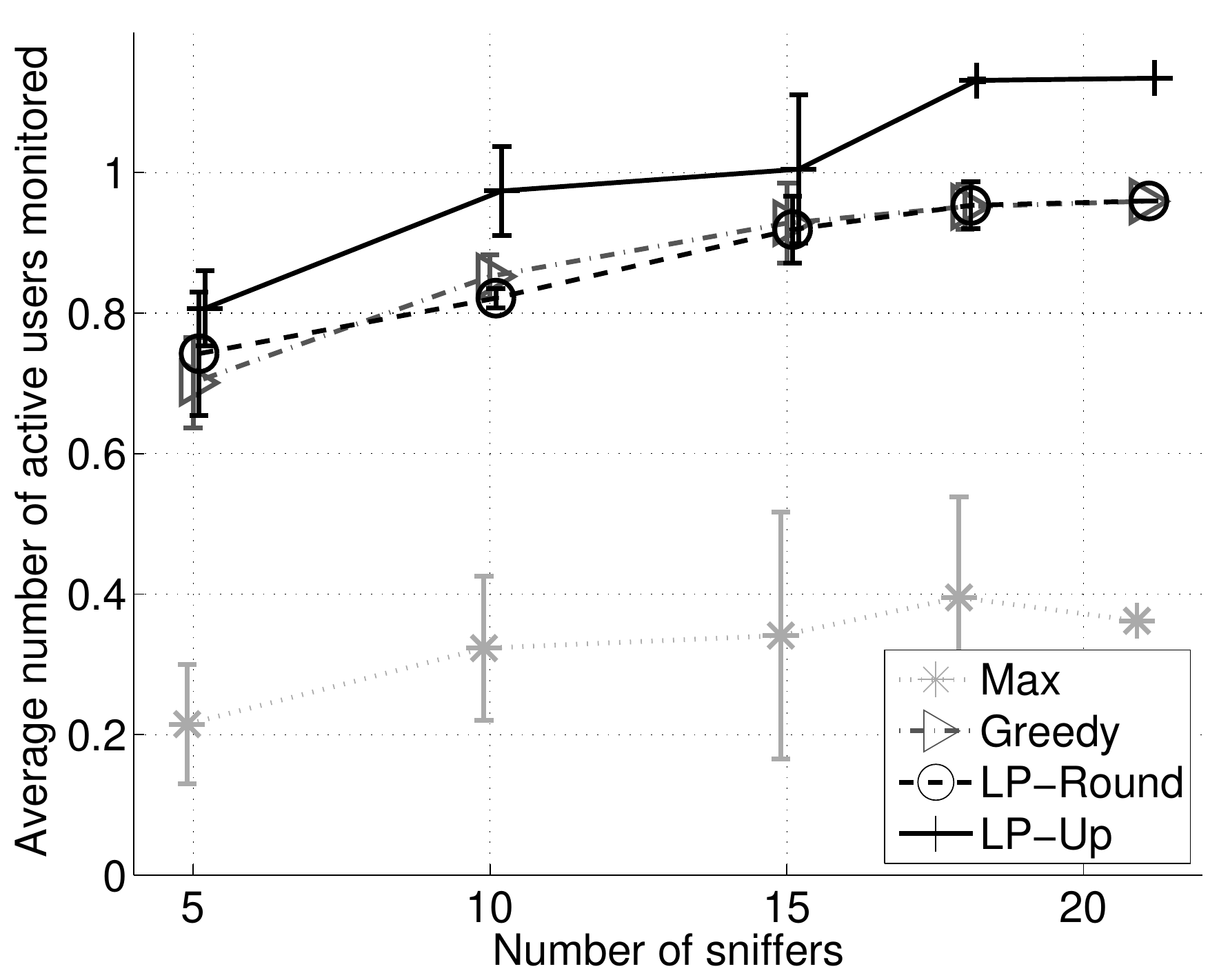} \\
(a) User-centric model &
(b) Quantized linear ICA &
(c) Binary ICA
\end{tabular}
\caption{QoM under the user-centric and sniffer-centric models with real WiFi
traces. In the user-centric model, the results of LP-round coincide with that
of the LP-Up. In some cases, the confidence interval is quite small and is thus
not observable in the figures.}
\label{fig:real_3}
\end{center}
\end{figure*}

\begin{figure}[t!]
\begin{center}
\includegraphics[width=3.6in]{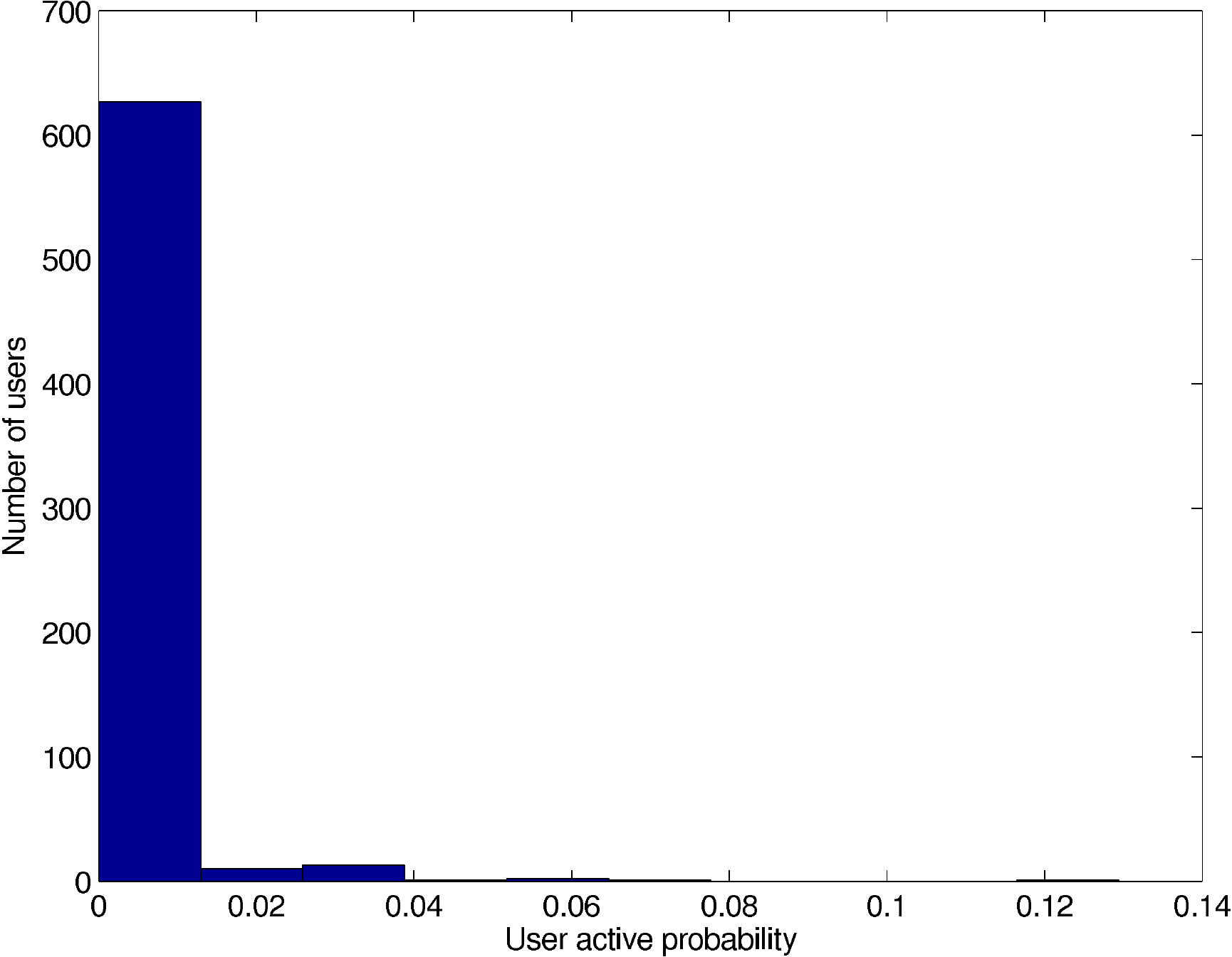}
\caption{Histogram of user active probability measured as the percentage of active 20$\mu$s slots. The average active probability is 0.0014.}
\label{fig:hist}
\end{center}
\end{figure}

Figure~\ref{fig:real_3} gives the average number of active users monitored
under the user-centric model, and under the models inferred by \qlica\ and \bica. The number of sniffers in
the experiments varies from 5 to 21 by including only traces from the
corresponding sniffers.  The number of channels is fixed at 3. Except for the
case with 21 sniffers, all data points are averages of 5 scenarios with
different sets of sniffers, chosen uniformly at random. Recall that the average
active probability is 0.0014. Thus, for the best channel assignment scenario,
the QoM on all channels is around 1.
In the user-centric case (Figure~\ref{fig:real_3}(a)), both
\greedy\ and LP-Round significantly outperform Max (by around 50\%).
Moreover, their performance is comparable
with LP-Up. As the number of
sniffers increases, the average number of users monitored increases but tends
to flatten out since most users have been monitored.

In the sniffer-centric case, similar trends can be observed when $\G$ and
$\MP(\yy)$ are inferred using \qlica\ and \bica\
(Figure~\ref{fig:real_3}(b)(c)). \bica\ outperforms \qlica\ in
general. However, there exists some performance gap in both cases due to the
loss of information, when compared with the user-centric model.  The real WiFi
traces, in contrast to the synthetic scenarios, contain a large number of
observations and many ``mice" users (users with very low active probability).
Most of the time, these users will be removed (since $p_i < \varepsilon$),
causing higher prediction errors in $\hat{\G}$ and $\hat{\pp}$.

\subsection{Comparison of \qlica\ and \bica\ under the sniffer-centric model}
To understand the performance difference of \qlica\ and \bica, in this section, we
provide a detailed comparison of the inferred $\G$ and $\MP(\yy)$ in both schemes.
\paragraph*{Performance metrics}
We denote by $\hat{\G}$ and $\hat{\pp}$ the inferred adjacency matrix
and the inferred active probability of users, respectively. Two metrics are introduced to measure the accuracy of the inferred quantities.
\begin{itemize}
\item {\bf Structure Error Ratio.}
This metric indicates how accurate the adjacency matrix is estimated.
It is defined by the Hamming distance between $\G$ and $\hat{\G}$
divided by the size of the matrix.
\beq
\begin{array}{lll}
\bar{H}(\G,\hat{\G}) & \stackrel{\Delta}{=} & \frac{1}{mn} \sum_{i=1}^{n} d^H(g_{i},\hat{g}_{i})
\end{array}
\eeq
Due to the possible difference in the number and the order of inferred independent
components in $\G$ and $\hat{\G}$, we need to perform the \textit{structure
matching process} before estimating $\bar{H}(\G,\hat{\G})$. Details of
the algorithm can be found in \cite{Nguyen2011TSP}.

\item {\bf Transmission Probability Error.}
The prediction error in the inferred transmission probability of
independent users is measured by the Kullback-Leibler divergence
between two probability distributions $\pp$ and $\hat{\pp}$. Let $\pp'$ and $\hat{\pp}'$
denotes the ``normalized'' $\pp$ and $\hat{\pp}$ ($\pp'_i = p_i\sum_{i=1}^{n}p_i$),
Transmission Probability Error is defined as below:
\beq
\begin{array}{lll}
\bar{P}(\pp',\hat{\pp}') & \stackrel{\Delta}{=} & \sum_{i=1}^{n}p_i\log(\frac{p_i'}{\hat{p_i}'})
\end{array}
\eeq
Intuitively, Transmission Probability Error gets larger as the predicted
probability distribution $\hat{\pp}$ is more deviated from the real distribution $\pp$.
\end{itemize}

\paragraph*{Results}
In this set of experiments, 10 sniffers and $n$ users are deployed on an
$1,000 \times 1,000$ square meter area, with $n$ varying from 5 to 20. Sniffers are placed
randomly on the area with the coverage radius set to 100 meters. In each run,
different sets of user locations are arbitrarily chosen. Only placements
satisfying the restriction that no two users are observed by a same set of sniffers
are included in the simulation. User transmission probability is selected
randomly in (0, 0.06].  All users and sniffers operate on a same wireless
channel since we are only interested in the accuracy of the inferred $\G$ and
$\MP(\yy)$.  The size of sample data $T = 10,000$.  Results are the average of
20 different runs.

\begin{figure*}[t]
\begin{center}
\begin{tabular}{cc}
\hspace{-0.35in}\includegraphics[width=4in]{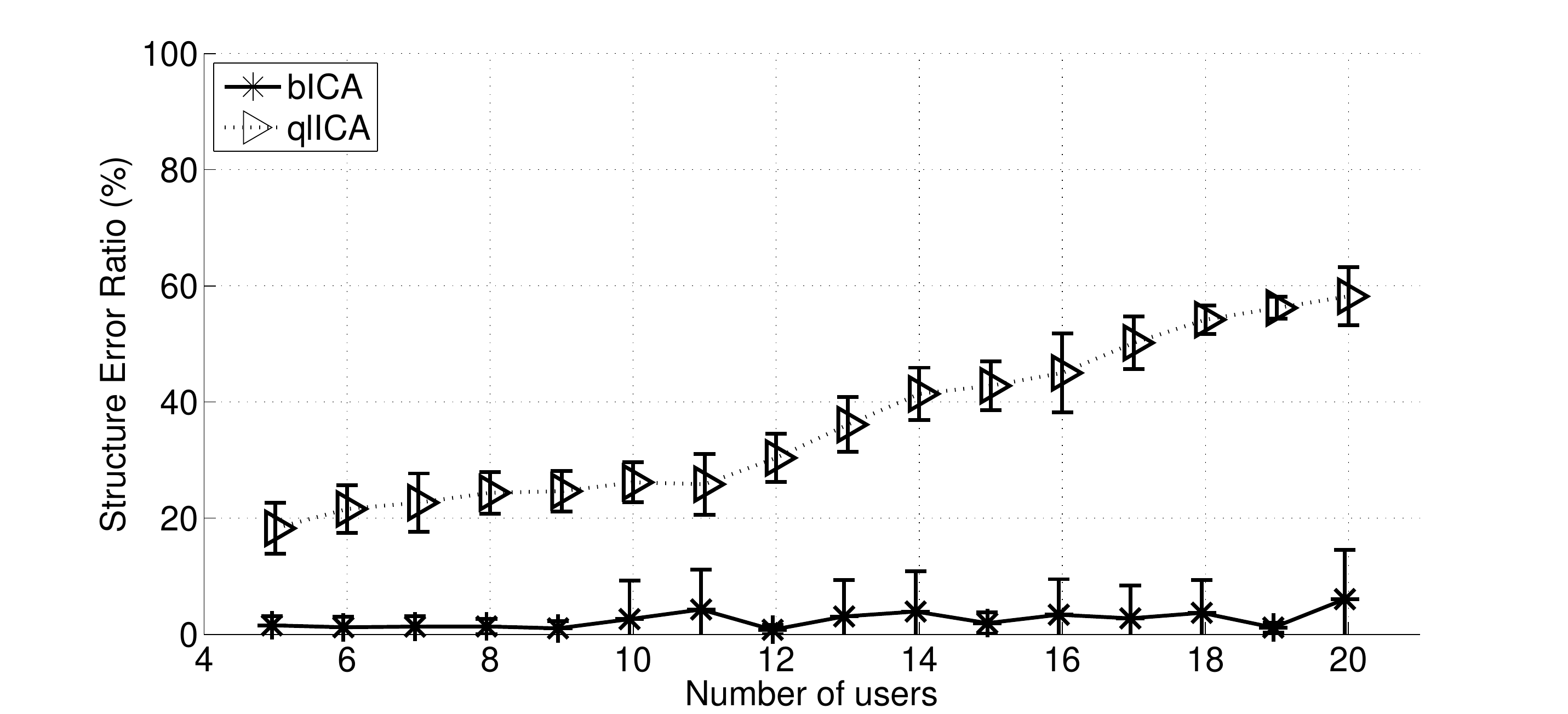} \hspace{-0.3in} & \hspace{-0.3in} \includegraphics[width=4in]{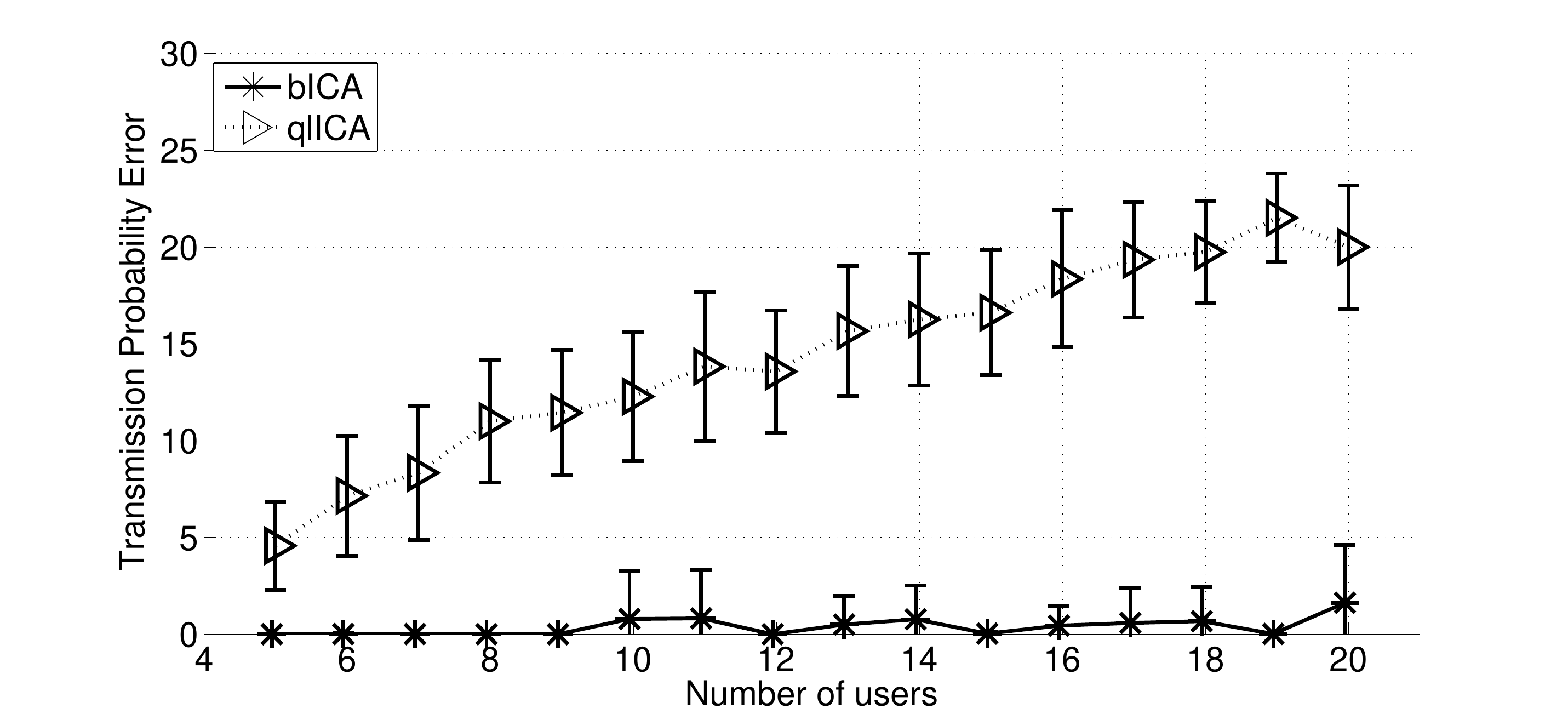}\\
\hspace{-0.35in}(a) Structure Error Ratio \hspace{-0.3in} & \hspace{-0.3in} (b) Transmission Probability Error \\
\end{tabular}
\caption{Accuracy validation result of two models \qlica\ and \bica. Result
is average of 20 runs with different initial seeds and symmetric error bars indicate
standard deviations.}
\label{fig:qlbICA}
\end{center}
\end{figure*}

From Figure~\ref{fig:qlbICA}, we see that \bica\ can achieve lower
prediction errors than \qlica\ on both $\G$ and $\MP(\yy)$. The former is
not very sensitive to the number of users, while the performance of \qlica\
degrades as the number of users increases.  This is somewhat expected as \qlica\ is
fundamentally a linear ICA method. Additionally, the estimation of $\MP(\yy)$ in
\qlica\ only utilizes first-order
statistics. In contrast, \bica\ is a joint procedure designed specifically
for binary data following disjunctive generation models.

\vspace{-0.1in}
\section{Discussion}
\label{sec:discuss}
In this section, we discuss several practical considerations in implementing
the proposed algorithms in real systems for wireless monitoring.

The primary focus of this work is sniffer-channel assignment given fixed
sniffer locations. Sniffer placement has been addressed in
\cite{shin09optimal}, which  assumes worse case loads in the network,
while sniffer-channel assignment can be made based on the actual
measured loads. In fact, both problems can be considered in a single
optimization framework if we generalize the sniffer placement problem to
decide online which set of sniffers should be turned on given budget constraints.

Implementation of sniffer-channel assignment should incorporate the learning
procedure proposed in \cite{AroraSZ11}. The time granularity of channel assignment should be sufficiently long to amortize the cost due to channel switching. To allow a consistent view of the channel at different locations, clock synchronization across multiple sniffers is needed. While clock synchronization can be performed offline using the frame traces collected~\cite{zheng09wiseranalyzer}, the accuracy of clock synchronization directly affects the inference accuracy of the ICA based methods in the sniffer-centric model. The choice of the slot of the binary measurements shall be made that takes into account the persistence of user transmission activities.

The channel assignment in its current form is computed in a centralized manner. This is reasonable since the sniffers are likely operated by a single administrative domain. An alternative distributed implementation has been considered in \cite{arora11gibbs} for the user-centric model based on the annealed Gibbs sampler. However, parameters of the distributed algorithm need to be properly tuned for fast convergence (and hence less message exchanges). From our understanding, the sniffer-centric model is not immediately amiable to distributed implementation.
\vspace{-0.1in}
\section{Conclusion}
\label{sec:conclusion}
In this paper, we formulated the problem of maximizing QoM in multi-channel
infrastructure wireless networks with different {\it a priori} knowledge.  Two
different models are considered, which differ by the amount (and type) of
information available to the sniffers.  We show that when complete information
of the underlying cover graph and access probabilities of users are
available, the problem is NP-hard, but can be approximated within a constant
factor.
When only binary information about the channel activities is available to the
sniffers, we propose two approaches (\qlica\ and \bica) so that one can map
the problem to the one where complete information is at hand using the
statistics of the sniffers' observations. We further conducted a detail
study comparing the performance of \qlica\ and \bica.
Finally, evaluations demonstrate the effectiveness of our proposed
inference methods and optimization techniques.

%
\vspace{-0.1in}
\section*{Acknowledgment}
The work of Nguyen and Zheng is funded in
part by the National Science Foundation (NSF) under award CNS-0832089, CNS-1117560.
Scalosub is partially supported by the CORNET consortium, sponsored by the
Magnet Program of Israel MOITAL.

%

\vspace{-0.1in}
\bibliographystyle{IEEEtran}


\vspace{-0.35in}
\begin{IEEEbiography}[{\includegraphics[width=1in,height=1.25in,clip,keepaspectratio]{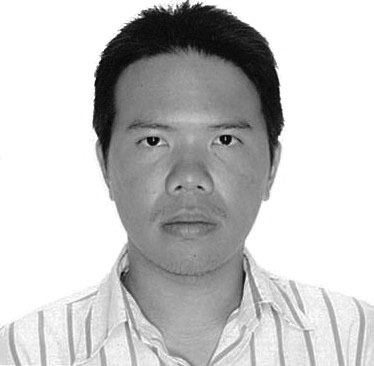}}]{Huy Nguyen}
received his B.S. degree in Computer Science from University of Science, Ho Chi Minh City, Vietnam, in 2006, and his M.E. degree in Electrical Engineering from Chonnam National University, Guangju, Korea, in 2009. Since 2009, he started pursuing his Ph.D. degree in the Department of Computer Science, University of Houston under the guidance of Prof. Rong Zheng. His research interests include wireless and sensor network management, information diffusion on social networks.
\end{IEEEbiography}
\vspace{-0.35in}
\begin{IEEEbiography}[{\includegraphics[width=1in,height=1.25in,clip,keepaspectratio]{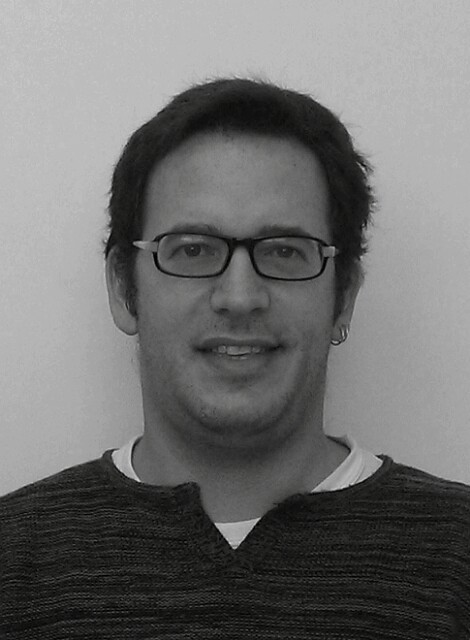}}]{Gabriel Scalosub}
received his B.Sc. in Mathematics and Philosophy from the Hebrew University of Jerusalem, Israel, in 1996. He received his M.Sc. and Ph.D. in Computer Science from the Technion - Israel Institute of Technology, Haifa, Israel, in 2002 and 2007, respectively. In 2008 he was a postdoctoral fellow at Tel-Aviv University, Israel, and in 2009 he was a postdoctoral fellow at the University of Toronto, Canada. In October 2009 Gabriel Scalosub joined the Department of Communication Systems Engineering at Ben Gurion University of the Negev, Israel. He served on various technical program committees, including Infocom, IWQoS, IFIP-Networking, ICCCN, and WCNC. His research focuses on theoretical algorithmic issues arising in various networking environments, including buffer management, scheduling, and wireless networks. He is also interested in broader aspects of combinatorial optimization, online algorithms, approximation algorithms, and algorithmic game theory.
\end{IEEEbiography}
\vspace{-0.35in}
\begin{IEEEbiography}[{\includegraphics[width=1in,height=1.25in,clip,keepaspectratio]{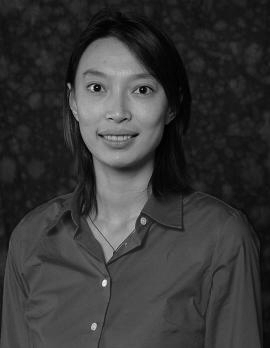}}]{Rong Zheng} (S'03-M'04-SM'10) received her Ph.D. degree from Dept. of Computer Science, University of Illinois at Urbana-Champaign and earned her M.E. and B.E. in Electrical Engineering from Tsinghua University, P.R. China. She is on the faculty of the Department of Computing and Software, McMaster University. She was with University of Houston between 2004 and 2012. Rong Zheng's research interests include network monitoring and diagnosis, cyber physical systems, and sequential learning and decision theory. She received the National Science Foundation CAREER Award in 2006. She serves on the technical program committees of leading networking conferences including INFOCOM, ICDCS, ICNP, etc. She served as a guest editor for EURASIP Journal on Advances in Signal Processing, Special issue on wireless location estimation and tracking, Elsevler's Computer Communications – Special Issue on Cyber Physical Systems; and Program co-chair of WASA'12 and CPSCom'12.
\end{IEEEbiography}

\vfill

\end{document}